\documentclass[letterpaper,twocolumn,10pt]{article}
\usepackage[10pt,nocopyright]{sigmin}

\usepackage{libertine}
\usepackage{libertinust1math} 
\usepackage{inconsolata}

\usepackage[breaklinks,colorlinks]{hyperref}
\usepackage{xcolor}
\definecolor{linkblue}{RGB}{0,40,200}
\hypersetup{%
  colorlinks=true,
  citecolor=linkblue,
  linkcolor=black,
  urlcolor=linkblue,
}

\input{glyphtounicode}
\usepackage[square,comma,numbers,sort&compress]{natbib}
\usepackage[T1]{fontenc}
\usepackage{xspace}
\usepackage{fancyvrb}
\usepackage{microtype}
\usepackage{fnpct}
\usepackage{graphicx}
\usepackage{amsmath,amssymb,amsthm}
\usepackage{aliascnt}
\usepackage{booktabs}
\usepackage[binary-units]{siunitx}

\usepackage{tikz}
\usepackage{tikzpeople}
\usetikzlibrary{shapes,arrows,automata,fit,backgrounds,calc}

\usepackage{underscore}
\usepackage{relsize}
\usepackage{bytefield}
\usepackage{enumitem}
\usepackage{subfigure}
\usepackage{rotating}
\usepackage{chngcntr}

\newcommand{\sys}{\textsf{AciKV}\xspace}
\newcommand{\ie}{{i.e.,}~}
\newcommand{\eg}{{e.g.,}~}

\newif\ifdraft\drafttrue
\newif\ifnotes\notestrue
\ifdraft\else\notesfalse\fi

\definecolor{xxxcolor}{rgb}{0.8,0,0}
\makeatletter
\long\def\XXX{\@ifnextchar[{\@XXX}{\@XXX[]}}
\long\def\@XXX[#1]{\@ifnextchar[{\@@@XXX{#1}}{\@@XXX{#1}}}
\ifnotes
\long\def\@@XXX#1{{\color{xxxcolor} XXX #1}\xspace}
\long\def\@@@XXX#1[#2]{{\color{xxxcolor} XXX (#1) #2}\xspace}
\else
\long\def\@@XXX#1{\ignorespaces}
\long\def\@@@XXX#1[#2]{\ignorespaces}
\fi
\makeatother

\fvset{fontsize=\footnotesize,xleftmargin=6pt}

\input{code/fmt}

\theoremstyle{definition}

\newaliascnt{lemma}{thm}

\aliascntresetthe{lemma}

\def\Snospace~{\S{}}

\numberwithin{equation}{section}

\counterwithout{equation}{section}

  {\begin{list}{$\bullet$}%
     {\setlength{\parsep}{0pt}%
      \setlength{\topsep}{0pt}%
      \setlength{\itemsep}{0pt}}}%
  {\end{list}}

\bibpunct[: ]{[}{]}{,}{n}{XXX}{XXX}

\newenvironment{denseitemize}{
\begin{itemize}[topsep=2pt, partopsep=0pt, leftmargin=1.5em]
  \setlength{\itemsep}{4pt}
  \setlength{\parskip}{0pt}
  \setlength{\parsep}{0pt}
}{\end{itemize}}

\def\headline#1{\hbox to \hsize{\hrulefill\quad\lower0.5ex\hbox{#1}\quad\hrulefill}}

\renewcommand{\FancyVerbFormatLine}[1]{%
\ifnum\ifnum\value{FancyVerbLine}=101 1\else\ifnum\value{FancyVerbLine}=116 1\else0\fi\fi=1%
\raisebox{0.5em}{\headline{\textbf{\textrm{#1}}}}
\else#1\fi}

\RecustomVerbatimEnvironment
{Verbatim}{Verbatim}
{xleftmargin=2pt,baselinestretch=0.95}

\definecolor{blue}{RGB}{0, 118, 186}

\newcommand{\sysget}{\emph{get(k)}\xspace}
\newcommand{\sysgetrange}{\emph{getrange(k1, k2)}\xspace}
\newcommand{\sysput}{\emph{put(k, v)}\xspace}
\newcommand{\sysdelete}{\emph{delete(k)}\xspace}
\newcommand{\sysbegin}{\emph{begin}\xspace}
\newcommand{\syscommit}{\emph{commit}\xspace}
\newcommand{\sysabort}{\emph{abort}\xspace}
\newcommand{\syspersist}{\emph{persist}\xspace}

\newcommand{\scflush}{\emph{flush}\xspace}

\newcommand{\fsync}{\texttt{fsync}\xspace}

\newcommand{\wsetnone}{\textsf{None}\xspace}
\newcommand{\wsetlist}{\textsf{List}\xspace}
\newcommand{\wsettree}{\textsf{Tree}\xspace}
\newcommand{\acid}{ACID\textsuperscript{--}\xspace}
\newcommand{\strunning}{\textsc{Running}\xspace}
\newcommand{\stobserving}{\textsc{Observing}\xspace}
\newcommand{\stcommitting}{\textsc{Committing}\xspace}
\newcommand{\staborted}{\textsc{Aborted}\xspace}
\newcommand{\stcommitted}{\textsc{Committed}\xspace}
\newcommand{\staccepting}{\textsc{Accepting}\xspace}
\newcommand{\stwaiting}{\textsc{Waiting}\xspace}
\newcommand{\stpersisting}{\textsc{Persisting}\xspace}

\newcommand{\leveldb}{\textsf{LevelDB}\xspace}
\newcommand{\rocksdb}{\textsf{RocksDB}\xspace}
\newcommand{\kyoto}{\textsf{Kyoto Cabinet}\xspace}
\newcommand{\bdb}{\textsf{Berkeley DB}\xspace}
\newcommand{\lmdb}{\textsf{LMDB}\xspace}
\newcommand{\sqlite}{\textsf{SQLite3}\xspace}

\usepackage[color=yellow,textsize=scriptsize]{todonotes}
\begin{document}

\title{\LARGE\bf Weakly Durable High-Performance Transactions}
\author{
    Yun-Sheng Chang\footnote{Work done while the author was affiliated with Academia Sinica.} \\ \emph{MIT CSAIL} \and 
    Yu-Fang Chen \qquad Hsiang-Shang Ko \\ \emph{Institute of Information Science, Academia Sinica, Taiwan}
}
\date{}

\maketitle

\begin{abstract}

%

Existing disk-based database systems largely fall into two categories---they either provide very high performance but few guarantees, or expose the transaction abstraction satisfying the full ACID guarantees at the cost of lower performance.
In this paper, we present an alternative that achieves the best of both worlds, namely good performance and transactional properties.
Our key observation is that, because of the frequent use of synchronization primitives, systems with strong durability can hardly utilize the extremely high parallelism granted by modern storage devices.
Thus, we explore the notion of weakly durable transactions, and discuss how to safely relax durability without compromising other transactional properties.
We present \sys, a transactional system whose design is centered around weak durability.
\sys exposes to users the normal transactional interface, but what sets it apart from others is a new ``persist'' primitive that decouples durability from commit.
\sys is a middle ground between systems that perform fast atomic operations, and ones that support transactions; this middle ground is useful as it provides similar performance to the former, while prevents isolation and consistency anomalies like the latter.
Our evaluation using the YCSB benchmark shows that \sys, under workloads that involve write requests, exhibits more than two orders of magnitude higher throughput than existing strongly durable systems.

\end{abstract}


\section{Introduction}
\label{sec:intro}

Existing disk-based database systems often sit on the two ends of the performance-guarantee spectrum.
At one end of the spectrum, we have systems that can perform operations in an extremely efficient manner, but often come with very few guarantees, for instance, the atomicity of a single operation.
At the other end of the spectrum, we have transactional database systems that guarantee the appealing atomicity, consistency, isolation, and durability (ACID) properties, but often have significantly lower performance compared with ones that have no support for transactions.
This state of affairs forces users to pick between good performance and useful abstraction.

\begin{figure}[t]
  \centering
  \subfigure{
    \scalebox{.9}{
      \begin{tikzpicture}
  \definecolor{colorsys}{HTML}{E65100}
  \draw[->, thick] (0, 0) -- (5, 0) node[below, anchor=north east] (guarantee) {\bf\normalsize Guarantee};
  \draw[->, thick] (0, 0) -- (0, 3) node[above left=0mm and 3mm, anchor=south west, rotate=0] (performance) {\bf\normalsize Performance};
  \node[fill, circle, minimum size=2mm, inner sep=0pt] (point-op) at (1.5, 2.2) {};
  \node[below=1mm of point-op] (op) {\normalsize Atomic operation};
  \node[star, minimum size=3mm, inner sep=0cm, fill=colorsys] (point-wd) at (3.3, 2.2) {};
  \node[above=1mm of point-wd] (wd) {\color{colorsys}\bf\normalsize \acid transaction};
  \node[fill, circle, minimum size=2mm, inner sep=0pt] (point-txn) at (4, 1) {};
  \node[below left=1mm and 8mm of point-txn.south, anchor=north] (txn) {\normalsize Full ACID transaction};
\end{tikzpicture}
    }
  }
  \subfigure{
    \includegraphics[width=3cm]{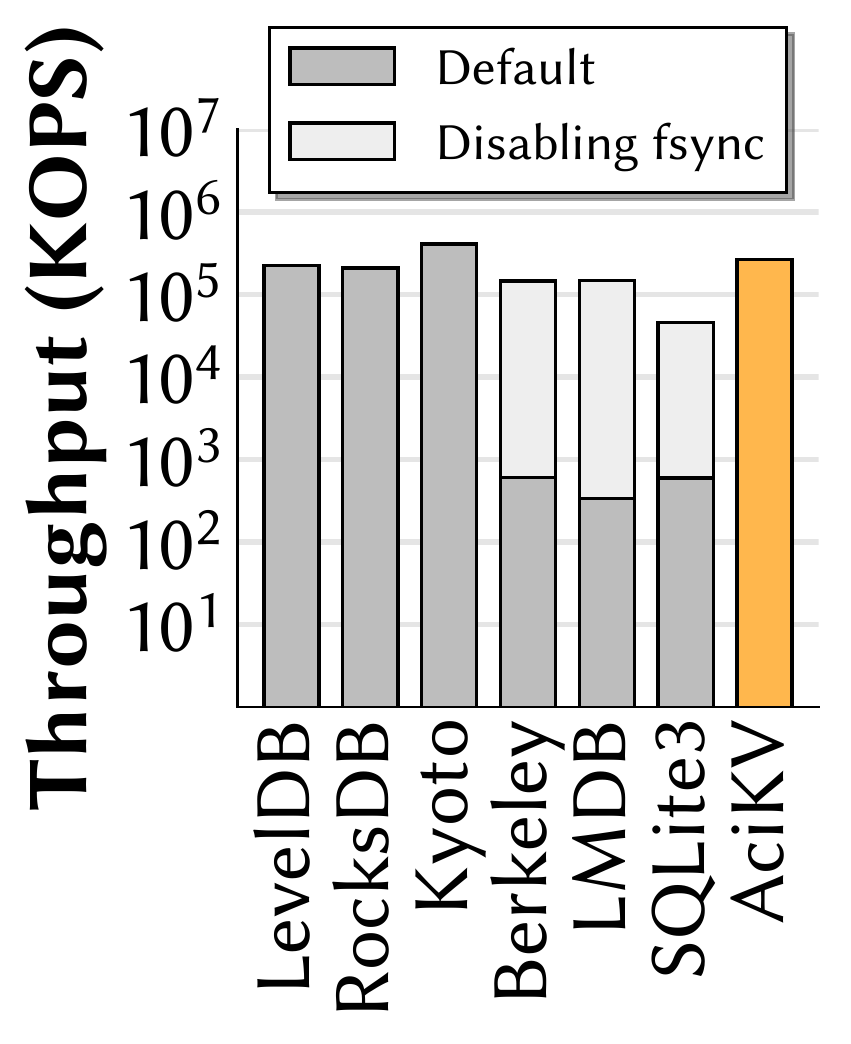}
  }
  \caption{\small
      Evaluation results of three non-transactional systems (\leveldb, \rocksdb, \kyoto), three transactional systems (\bdb, \lmdb, \sqlite), and this work (\sys) under a 50R-50W workload.
    See \autoref{subsec:exp-setup} for the detailed experimental setup.
  }
  \label{fig:overview}
\end{figure}
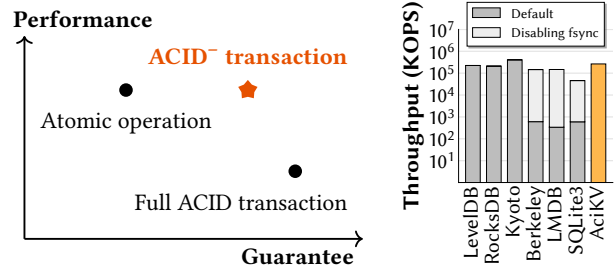

We argue that this unsatisfactory situation can be primarily attributed to the pursuit of \emph{strong durability}, which induces high synchronization overhead.
To validate this argument, we ran an experiment on some popular transactional systems in two modes: a default strongly durable mode, and a weakly durable mode that disables the issuing of \fsync \footnote{Throughout this paper, we use \fsync to denote file system synchronization primitives including \texttt{fdatasync} and \texttt{sync}.} to the underlying file system.
As shown in \autoref{fig:overview}, all systems in the weakly durable mode performs at least two orders of magnitude faster than when they are in the default strongly durable mode.
A similar experiment conducted recently at the level of file systems and of disks~\cite{won18:barrierfs} also aligns with our result.
This analysis motivates us to explore the notion of \emph{weakly durable transactions}, which aims to perform comparably to systems capable of carrying out fast operations, while providing the useful transaction abstraction.

Another motivation for this work is hardware trend.
Modern storage devices, such as flash-based solid-state drives (SSDs) and emerging non-volatile memory (NVM) technology, are often structured in a highly parallel way.
While they promise to provide tremendous throughput, strongly durable systems can hardly utilize the granted parallelism because of the frequent use of synchronization primitives.

A common technique that can also alleviate the synchronization overhead is \emph{group commit}, which groups multiple transactions into one batch, and delays issuing a synchronization primitive (\eg an \fsync) until the end of each batch.
The problem with group commit, however, is an inherent trade-off between throughput and latency---a large batch increases throughput but suffers from higher latency, and a small batch has a lower latency but also a lower throughput.
In \autoref{subsec:group-commit}, we experimentally evaluate the two techniques, weak durability and group commit, and discuss their differences in detail.

At this point, one might ask: Isn't disabling \fsync good enough?
The answer is no, as simply disabling \fsync can easily lead to \emph{internal inconsistency}.
This means that the consistency invariants (\eg an allocated data block should always be referenced by some metadata) enforced internally by the system may no longer hold true after a full system crash (\eg a power failure).
A mild consequence is resource leaks, and a more severe one is unusable corrupted databases~\cite{zheng14:torturing-db}.
The reason internal inconsistency may arise is that file system operations may be reordered on a crash~\cite{bornholt16:ccmodel, pillai14:ccbug}, and without \fsync, the database system cannot enforce certain write orderings for crash safety.
Thus, the LMDB developer faithfully warns its users about the risk of enabling such option~\cite{lmdbcorrupt}:
\begin{quote}
``MDB_NOSYNC \ldots This optimization means a system crash can corrupt the database or lose the last transactions if buffers are not yet flushed to disk. \ldots However, if the filesystem preserves write order and the MDB_WRITEMAP flag is not used, transactions exhibit ACI (atomicity, consistency, isolation) properties and only lose D (durability).''
\end{quote}
SQLite3 also gives their users a similar warning~\cite{sqlite3corrupt}.
But there is a more convoluted issue.
In the context of database transactions, consistency (\ie C in ACID) is often specified with a set of \emph{integrity constraints}.
For instance, one might require the account balance to be non-negative, or the user identifier to be unique.
Without any restriction on weak durability, it is possible for a crash to create \emph{external inconsistency}; that is, a database user may observe a violation of the integrity constraints after a crash.
In \autoref{subsec:txn-sys}, we give an example of external inconsistency induced by weak durability, and elaborate more on its cause.

Naturally, our exploration of weakly durable transactions starts with a theoretical analysis on how to \emph{safely} relax durability to avoid inconsistency; \autoref{subsec:crash-consistency} shows that a mild adaptation of the standard serializability theory~\cite{bernstein87:ccr} is sufficient for this purpose.
The refined weak durability, along with other transactional properties, gives rise to what we call the \acid properties.
\autoref{tbl:api} shows an interface of weakly durable transactional systems.
\begin{table}[h]
  \small\centering
  \begin{tabular}{@{} lr @{}}
	\toprule
	\textbf{Category} & \textbf{Primitive} \\
	\midrule
	Data & \sysget, \sysgetrange, \sysput, \sysdelete \\
	Transaction & \sysbegin, \syscommit, \sysabort \\
	Durability & \syspersist \\
	\bottomrule
  \end{tabular}
  \caption{An interface of weakly durable transactional systems.}
  \label{tbl:api}
\end{table}
The crux of weak durability is manifested in the new \syspersist primitive, which decouples durability from \syscommit: committed transactions are not guaranteed to survive across crashes; only when they encounter a \syspersist do they become persistent.
However, users can rest assured that no inconsistency will arise because of crashes.



As a demonstration, we built \sys, which implements the interface in \autoref{tbl:api} and guarantees \acid.
While the main benefit of weak durability is a significant reduction in synchronization overhead, we believe the \emph{permissiveness} of weak durability also implies a simpler and more efficient system design.
Thus, we built \sys from scratch using weak durability as its ``first-class'' property (rather than building it to ensure strong durability, and trying to relax durability later on).
In \autoref{subsec:shadow}, we introduce the recovery mechanism of \sys, including an efficient client-server synchronization protocol and a shadow paging technique \cite{lorie77:shadow}, which together allow \sys to easily create a consistent snapshot in a crash-safe manner.
Compared with the more commonly used ARIES-style~\cite{mohan92:aries} write-ahead logging (WAL) that often has to depend on other layers such as buffers, or to assume a certain locking protocol, \sys's recovery mechanism is much simpler, in that it \emph{fully decouples crash safety from other layers}.
We also discuss why the conventional wisdom that shadow paging is not suitable for concurrent writes~\cite{gray81:systemr-recovery} may not be the case for weak durability.
In \autoref{subsec:index}, we describe \sys's latch-free two-level index design that fits well with the \emph{batch-processing} nature of weak durability.
The index consists of an in-memory concurrent skip list and an on-disk B+-tree.
The skip list is employed for absorbing insertions, so that the B+-tree structure is guaranteed to remain the same until the next \syspersist, at which point the two data structures are merged.
This means that no index latches are required.
Moreover, unlike other multi-level index structures such as the LSM-tree~\cite{oneil96:lsm-tree}, \sys does not need an extra mechanism to protect data in the skip list, as the data are changes made after the last \syspersist.

Using the YCSB benchmark suite~\cite{cooper10:ycsb}, we compare \sys with six popular disk-based database systems, three transactional and three non-transactional.
Evaluation results show that, under workloads that involve writes, the throughput of \sys can be two or even three orders of magnitude higher than that of the transactional systems, and is on par with that of the non-transactional ones.
Further, to understand the benefit of building \sys around weak durability, we also compare \sys with the transactional systems running in their weakly durable mode.
Results show that \sys is the only system that consistently scales across a wide range of workloads.

\paragraph{\bf Limitations of \sys.}
\begin{denseitemize}
\item
\sys is a \emph{storage engine}, whose main responsibilities are concurrency control and crash recovery.
Thus, its interface supports only the generic byte-array data type and a record-at-a-time data access method, and has no mechanisms for specifying integrity constraints.
System designers can build their own data model (\eg SQL) on top of \sys.
\item
\sys lacks features commonly seen in database systems, such as compression and replication.
We believe that \sys's design and properties are compatible with these features, so adding them to \sys should be feasible.
\end{denseitemize}

\paragraph{\bf Contributions of this work.}
\begin{denseitemize}
\item
    We introduce the notion of weakly durable transactions, which aims to provide the useful transaction abstraction with enough guarantees, namely \acid, while maintaining good performance.
\item
    We present \sys, a transactional key-value store whose design is centered around weakly durable transactions.
\item
    We evaluate \sys by comparing it with six popular disk-based database systems using the YCSB benchmark.

\end{denseitemize}


\section{From ACID to \acid}
\label{sec:weak-durability}

Weakly durable transactional systems guarantee \acid, whose semantics follows that of standard ACID except for Durability, which is changed to guarantee that only those committed transactions made persistent by the \syspersist primitive (shown in \autoref{tbl:api}) will survive across crashes.
In this section, we first review ``ad-hoc'' weak durability seen in existing transactional systems~(\autoref{subsec:txn-sys}); in particular, we look at a simple example of inconsistency that may arise, showing that this kind of weak durability is potentially unsafe.
Then, we present our theoretical analysis on how to safely relax durability to maintain consistency~(\autoref{subsec:crash-consistency}).

\subsection{Ad-hoc weak durability}
\label{subsec:txn-sys}

ACID has been the gold standard of transactional systems for decades.
We briefly recap its definition here.
Atomicity says that the effect of each transaction should appear as a whole to other transactions.
Consistency ensures that each transaction always observes a consistent state satisfying all the given integrity constraints; also, on a crash, the recovered state should be consistent.
Isolation creates the illusion that each transaction is the sole owner of the database, even though multiple transactions may be executed concurrently.
Durability guarantees that the effect of committed transactions will always survive across crashes.
By promising these properties, transactions can often greatly relieve the programming burden of users.

But supporting the strongest form of ACID is expensive.
Many transactional systems therefore provide options to relax some of the guarantees, so that users can ``pay'' only for the guarantees they actually need.
For instance, many systems can run transactions at a lower isolation level.
In this work, however, we will discuss only systems with atomicity and serializability, the highest isolation level (and focus on the interplay between consistency and durability).
Some transactional systems also have options to relax durability, often achieved by not issuing slow synchronization primitives such as \fsync.
As shown in \autoref{fig:overview}, these options have the potential to significantly improve the performance.
To distinguish this kind of weak durability from the one we propose in this work, we refer to the former as \emph{ad-hoc weak durability}.

The downside of ad-hoc weak durability, however, is the risk of internal and external inconsistency when encountering a crash.
As we have covered internal inconsistency in \autoref{sec:intro}, here we focus on the external one, that is, violations of integrity constraints.
The reason ad-hoc weak durability is subject to inconsistency is that it does not specify \emph{which transactions are allowed to be discarded on a crash}.
Below we illustrate the issue with an example.
Suppose we have a database consisting of two integer objects $x$ and $y$, and an integrity constraint $x < y$.
Consider the following serial history:
\begin{equation*}
\setlength{\abovedisplayskip}{1ex}
\setlength{\belowdisplayskip}{1ex}
r_1(x, 0) r_1(y, 1) w_1(y, 2) c_1 r_2(x, 0) r_2(y, 2) w_2(x, 1) c_2
\end{equation*}
In the history, transaction T1 first observes $(x, y) = (0, 1)$, updates $y$ to $2$ without violating the constraint $x < y$, and commits; transaction T2 then observes $(x, y) = (0, 2)$, so it is also safe for T2 to update $x$ to $1$.
Assume a crash occurs after T2 commits.
While both transactions preserve the constraint, inconsistency can still arise in the post-crash state when we naively allow the weakly durable system to discard \emph{any} transaction on a crash---if the system chose to discard the effect of T1, the resulting state would be $(x, y) = (1, 1)$, violating $x < y$.

The previous example suggests that we need a refined version of weak durability to safely relax durability.
Before diving into our theoretical analysis, we begin with some intuition.
Note that there are four possible scenarios after the crash: (i) only T1, (ii) only T2, (iii) both T1 and T2, (iv) none of them survives across the crash, but only the second case can give rise to inconsistency.
The root cause of the problem is that \emph{T2 depends on T1's effect}, or more concretely, T2 observes T1's update of $y$ from $1$ to $2$.
Naturally, we should refine our definition of weak durability to incorporate the \emph{dependencies} between transactions.

Note that we by no means claim that ad-hoc weak durability \emph{will} create inconsistency, and indeed, a particular system may be free from inconsistency due to certain implementation artifacts.
The point being made is that we should justify \emph{why} a weakly durable system is not subject to inconsistency.


\subsection{Consistency of weakly durable systems}
\label{subsec:crash-consistency}

To make sure theoretically that consistency will not be lost if we adopt weak durability, we analyze the standard crash consistency argument for database systems with (strong) durability, serializability, and consistency preservation by all transactions~\cite[Section~6.2]{bernstein87:ccr}.
We will see that, with some mild adaptation, essentially the same argument also works for weakly durable systems.

The high-level argument is fairly straightforward.
Let modifications to a database be represented as a history~$H$ of read, write, commit, or abort operations performed by transactions.
When the database system crashes, (strong) durability guarantees that the recovered database retains exactly the effects of those operations performed by \emph{committed} transactions in~$H$.
This retained part of~$H$ is called the \emph{committed projection} of~$H$ and denoted by $C(H)$.
Serializability then guarantees that the recovered database is the same as the result of executing all the transactions in $C(H)$ in some serial order.
Finally, since each of the transactions preserves consistency, the recovered database must be consistent.

The durability guarantee is more involved than it seems, as it may not even be theoretically possible to retain exactly the effects of the operations in $C(H)$ due to the \emph{dependencies} among the operations in~$H$ (which form a partial order, so in general $H$~may be an acyclic directed graph rather than a sequence).
Here we do not need to discuss specific forms of dependency (\eg read-after-write dependency), and only need to assume that if an operation~$\mathit{op}$ depends on another operation~$\mathit{op}'$, then the effect of~$\mathit{op}'$ must happen before $\mathit{op}$~can take effect.
(For example, $\mathit{op}$ and $\mathit{op}'$ may be ``reading an object~$x$'' and ``writing to~$x$'', and the effect of $\mathit{op}$, namely getting a particular value~$v$ of~$x$, cannot happen if $\mathit{op}'$ does not take effect because $v$~has to be written by $\mathit{op}'$ first.)
Dependencies may prevent the database system from retaining exactly the effects of the operations in $C(H)$, because an operation in $C(H)$ may depend on some other operation not in $C(H)$ and could not take effect.
Therefore, to make durability at least theoretically possible (before we can move on to think about how to implement durability), we should ensure that $C(H)$ is a \emph{prefix} (or a ``downward closed'' subset) of~$H$, meaning that for every operation in $C(H)$, all the operations it depends on are in $C(H)$ as well.
This can be ensured by committing transactions with some care, and one way is to enforce the following \emph{prefix preservation} property (which is comparable to ``recoverability''~\cite[Section~2.4]{bernstein87:ccr}): whenever an operation of a transaction~$T$ depends on an operation of another transaction~$T'$, if $T$~is committed, then $T'$~must be committed before $T$~is.
This implies that $C(H)$ is a prefix of~$H$: every operation~$\mathit{op}$ in $C(H)$ belongs to a committed transaction, so for any operation~$\mathit{op}'$ in~$H$ on which $\mathit{op}$~depends, the transaction containing $\mathit{op}'$ must be committed, and hence $\mathit{op}'$ is in $C(H)$ as well.

For a database system with weak (instead of strong) durability, we can reuse essentially the same crash consistency argument, except that $H$~can now include \syspersist operations, and what the system retains after a crash is the effect of the \emph{persistently committed projection} $\mathit{PC}(H)$, which consists of the operations of the transactions committed before any \syspersist operation in~$H$.
Like in the case of strong durability, we should ensure that $\mathit{PC}(H)$ is a prefix of~$H$ so that weak durability is at least theoretically possible.
For this, the same prefix preservation property is sufficient: given an operation $\mathit{op}$ of a persistently committed transaction~$T$, any operation $\mathit{op}'$ on which $\mathit{op}$ depends must belong to a transaction~$T'$ which, by prefix preservation, is committed before $T$~is, and hence also persistently committed (due to the same \syspersist operation that makes~$T$ persistently committed).

The above analysis shows that, practically, it is possible to reuse some existing scheduling mechanism with prefix preservation and serializability to implement weakly durable systems that guarantee crash consistency.
Indeed this is what we do for our system \sys.


\section{\sys Design}
\label{sec:design}

\begin{figure}[t]
  \centering
  \scalebox{.8}{\begin{tikzpicture}
  \tikzstyle{slnode} = [draw=black, fill=gray!10, minimum size=2ex, rectangle, inner sep=0pt]
  \tikzset{
    pics/sl/.style n args={3}{
      code = {
        \node[slnode, right=3mm of #1] (#2) {#3};
        \draw[->, >=stealth] (#1) -- (#2);
      }
    }
  }
  \tikzstyle{treenode} = [
    minimum width=2ex, minimum height=2ex,
    rectangle, inner sep=0pt]
  \tikzset{
    pics/btree/.style n args={4}{
      code = {
        \node[treenode] (#1y) {\bf #3};
        \node[treenode, left=0pt of #1y] (#1x) {\bf #2};
        \node[treenode, right=0pt of #1y] (#1z) {\bf #4};
        \begin{scope}[on background layer]
          \fill[blue!10, draw=black] (#1x.north west) rectangle (#1z.south east);
        \end{scope}
      }
    },
    pics/btreestb/.style n args={4}{
      code = {
        \node[treenode] (#1y) {\bf #3};
        \node[treenode, left=0pt of #1y] (#1x) {\bf #2};
        \node[treenode, right=0pt of #1y] (#1z) {\bf #4};
        \begin{scope}[on background layer]
          \fill[green!10, draw=black] (#1x.north west) rectangle (#1z.south east);
        \end{scope}
      }
    },
  }
  \tikzset{
    tblnode/.style = {
      minimum height=2mm, minimum width=7mm, inner sep=0pt, draw=black
    },
    pics/tbl/.style n args={4}{
      code = {
        \footnotesize
        \node[tblnode, fill=#4, anchor=north east] (#1r0c0) {};
        \node[right=-\pgflinewidth of #1r0c0, tblnode, fill=#4] (#1r0c1) {};
        \node[above=1mm of #1r0c0, anchor=base] {#2};
        \node[above=1mm of #1r0c1, anchor=base] {#3};
        \node[below=-\pgflinewidth of #1r0c0, tblnode, fill=#4] (#1r1c0) {};
        \node[below=-\pgflinewidth of #1r0c1, tblnode, fill=#4] (#1r1c1) {};
        \node[below=-\pgflinewidth of #1r1c0, tblnode, fill=#4] (#1r2c0) {...};
        \node[below=-\pgflinewidth of #1r1c1, tblnode, fill=#4] (#1r2c1) {...};
      }
    }
  }
  \tikzset{
    layertext/.style = {
      minimum height=6ex, anchor=east, fill=white
    },
    layer/.style = {
      rectangle, dashed, draw=black, minimum width=77mm, minimum height=11mm
    },
  }
  \tikzstyle{arrowhead} = [draw=black, fill=white, circle, minimum size=4pt, inner sep=0pt]
  \node[layer] (fs) {};
  \node[layertext] (fstext) at ([xshift=-1em]fs.north east) {\bf File system};
  \node[anchor=center, rectangle, minimum width=2.4cm, minimum height=3ex, draw=black, fill=black!10] at (fs.center) (file) {};
  \node[left=0pt of file] {database file};

  \node[above=3mm of fs, layer] (shadow) {};
  \node[layertext] at ([xshift=-1em]shadow.north east) (shadowtext) {\textbf{Shadow} (\autoref{subsec:shadow})};
  \node[anchor=center, rectangle, minimum width=1.8cm, minimum height=3ex, draw=black, fill=blue!10] at ({file.west |- shadow.center}) (cur) {};
  \node[left=0pt of cur] {current file};
  \node[anchor=center, rectangle, minimum width=1.8cm, minimum height=3ex, draw=black, fill=green!10] at ({file.east |- shadow.center}) (stb) {};
  \node[right=0pt of stb] {stable file};
  \node[arrowhead] at (shadow.south) (shadow-circle) {};
  \draw[->] (shadow-circle) -- (fs);

  \node[above=3mm of shadow, layer, minimum height=27mm] (idx) {};
  \node[layertext] at ([xshift=-1em]idx.north east) (idxtext) {\textbf{Index} (\autoref{subsec:index})};
  \draw pic[below left=5mm and 18mm of idx.north] {btree={root}{.}{.}{}};
  \draw pic[below left=3mm and 5mm of rooty.south] {btree={b1}{.}{}{}};
  \draw pic[below right=3mm and 5mm of rooty.south] {btree={b2}{.}{.}{}};
  \draw pic[below left=3mm and 5mm of b1y.south] {btree={b3}{.}{.}{}};
  \draw pic[below right=3mm and 5mm of b1y.south] {btree={b4}{.}{}{}};
  \draw pic[below right=3mm and 5mm of b2y.south] {btree={b5}{.}{.}{}};
  \draw[->, >=stealth] (rootx.south) -- (b1y.north);
  \draw[->, >=stealth] (rooty.south) -- (b2y.north);
  \draw[->, >=stealth] (b1x.south) -- (b3y.north);
  \draw[->, >=stealth] (b2x.south) -- (b4y.north);
  \draw[->, >=stealth] (b2y.south) -- (b5y.north);
  \node[fit=(b3x)(b5z), inner sep=0pt] (leafgroup) {};
  \draw[->, >=stealth] (b5x) -- (b4z);
  \draw[->, >=stealth] (b4x) -- (b3z);
  \node[below=0pt of b4y] {current B+-tree};

  \node[right=0mm of b5z] (h0) {};
  \draw pic {sl={h0}{n0}{}};
  \draw pic {sl={n0}{n1}{}};
  \draw pic {sl={n1}{n2}{}};
  \draw pic {sl={n2}{n3}{}};
  \draw pic {sl={n3}{n4}{}};
  \draw pic {sl={n4}{n5}{}};
  \node[slnode, above=2mm of n2] (n6) {};
  \draw[->, >=stealth] (n6) -- (n2);
  \node[slnode, above=2mm of n4] (n7) {};
  \draw[->, >=stealth] (n7) -- (n4);
  \draw[->, >=stealth] (n6) -- (n7);
  \node[slnode, above=2mm of n6] (n8) {};
  \draw[->, >=stealth] (n8) -- (n6);
  \node (h1) at ({h0 |- n6}) {};
  \draw[->, >=stealth] (h1) -- (n6);
  \node (h2) at ({h0 |- n8}) {};
  \draw[->, >=stealth] (h2) -- (n8);
  \node[below=0pt of n2.south] {in-memory skip list};

  \node[arrowhead] at (idx.south) (idx-circle) {};
  \draw[->] (idx-circle) -- (shadow);

  \coordinate (lock-bottomleft) at ([xshift=3mm]fs.south east);
  \coordinate (lock-topright) at ([xshift=25mm]idx.north east);
  \coordinate (lock-bottomright) at ([xshift=25mm]fs.south east);
  \coordinate (lock-north) at ([xshift=14mm]idx.north east);
  \draw[dashed] (lock-bottomleft) rectangle (lock-topright);
  \node[layertext] at ([xshift=-1em]lock-bottomright) {\textbf{Lock} (\autoref{subsec:s2pl})};
  \path (lock-bottomleft) --
    pic[midway, above=10mm] {tbl={reclock}{key}{lock}{red!10}}
    pic[midway, below=10mm] {tbl={gaplock}{key}{lock}{red!10}}
    (lock-topright);
  \node[below=0pt of reclockr2c0.south east, fill=white] {record lock table};
  \node[below=0pt of gaplockr2c0.south east, fill=white] {gap lock table};

  \node[above=3mm of idx.north west, anchor=south west, layer, minimum width=102mm, minimum height=10ex] (server) {};
  \node[layertext] at ([xshift=-1em]server.north east) (servertext) {\textbf{Server} (\autoref{subsec:top-op})};
  \draw pic[below=5mm of server.north] {tbl={kv}{key}{val}{yellow!10}};
  \node[left=0pt of kvr1c0] {key-value mapping};
  \node[arrowhead] at ({lock-north |- server.south}) (server-circle-right) {};
  \draw[->] (server-circle-right) -- (lock-north);
  \node[arrowhead] at ({idx.north |- server.south}) (server-circle-left) {};
  \draw[->] (server-circle-left) -- (idx);


\end{tikzpicture}}
  \caption{\small
Overview of the \sys design.
Each layer implements its own state on top of its underlying layers.
\protect\begin{tikzpicture}
  \protect\node[draw=black, fill=white, circle, minimum size=3.2pt, inner sep=0pt] at (0, 2mm) (artail) {};
  \protect\draw[->] (artail) -- (0, 0);
\protect\end{tikzpicture}
: use.
  }
  \label{fig:design-overview}
\end{figure}
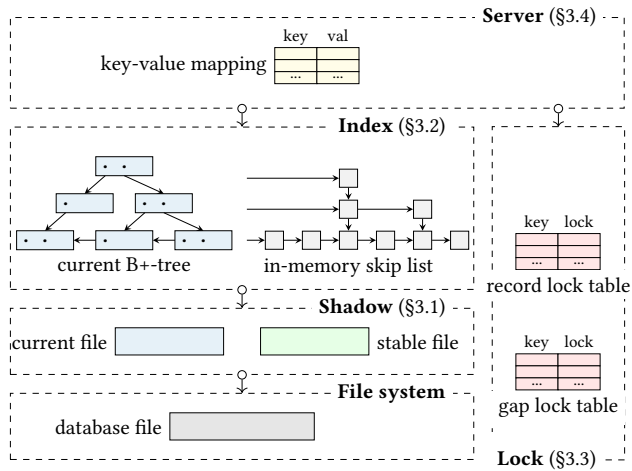

In this section, we present the design of \sys.
\autoref{fig:design-overview} shows its main layers.
Starting from a database file, \sys builds on top of it a \emph{shadow paging} mechanism~(\autoref{subsec:shadow}) to ensure crash safety.
More concretely, the shadow guarantees to revert the database to a consistent state after a system crash.
Similarly to all database systems, \sys uses an indexing mechanism~(\autoref{subsec:index}) to speed up point and range queries. 
Its index has two parts: an on-disk B+-tree and an in-memory skip list. 
On receiving a \emph{persist}, \sys will merge the skip list into the B+-tree and persistently create a consistent snapshot with the shadow for durability.
To ensure serializability of database transactions, \sys adopts the strong strict two-phase locking (SS2PL) protocol~(\autoref{subsec:s2pl}) on records and key intervals.
On top of the indexing mechanism and SS2PL, \sys provides standard operations (in addition to \syspersist) of key-value stores for its users~(\autoref{subsec:top-op}).
For simplicity, we omit layers (\eg the buffer layer) that are standard and whose absence does not obscure the exposition of \sys's behavior.

\subsection{Shadow paging}
\label{subsec:shadow}

Dealing with crashes is hard, as one has to thoroughly consider every possible post-crash states and carefully order certain writes to prevent crash-safety bugs.
To ensure crash safety, \sys relies on the \emph{shadow paging} (or simply shadow) technique~\cite{hitz94:wafl, rodeh08:btree-shadow, gray81:systemr-recovery, lorie77:shadow, chang20:scftl, rodeh13:btrfs}.
We choose to implement a modern and efficient one presented in~\cite{chang20:scftl}.
Below we briefly describe its design; for more details (in particular its correctness), we refer readers to the original paper.

At the core of the shadow is an in-memory page table, mapping logical addresses (\ie the ones upper layers observe) to physical addresses (\ie the ones being used to index the underlying database file).
Modifications are made in a out-of-place manner: On receiving a \emph{write} request, the shadow assigns a free physical page to hold the data, and updates the corresponding entry in the page table to point to the assigned page. 
Notice that due to the out-of-place update, the old data remain intact on the database file; this is crucial as the recovery procedure may need the old data to revert the database to a consistent state.

On receiving a \scflush request, the page table will be stored on the database file for durability, in the form of table differences (\ie the deltas), and occasionally when running out of space to accommodate the deltas, in the form of a full table image.
We call the full table image plus the deltas the \emph{stable table}. 
To ensure write ordering and durability, \texttt{fsync}s are issued as needed.
The garbage collector is also designed in a way that will not remove data pointed to by the stable table.

On recovery, the full table image is used as the base image, and the deltas are applied sequentially to rebuild the stable table.
In addition, since all data pointed to by the stable table are still there, the procedure can safely recover the file state to the time when the last \scflush command succeeds.

The shadow has a simple specification, consisting of two files \emph{current} and \emph{stable}.
The upper layer can \emph{read} and \emph{write} only the current one, and crash-atomically store a \emph{snapshot} of the current one into the stable one with the \scflush command.
On crashes, there is no warranty about the contents of the current file, but the stable file is guaranteed to remain intact so that during \emph{crash recovery}, the shadow can safely bring the snapshot back from the stable to the current one.
On top of the shadow, system designers can implement arbitrary data structures; for instance, \sys implements a B+-tree.


With the shadow at our disposal, ensuring crash safety is effectively reduced to \emph{creating a consistent snapshot}.
This means that we can update the stable storage as if crashes are absent, which frees us from reasoning about the intricate crash behavior and thus greatly simplify the design of upper layers.

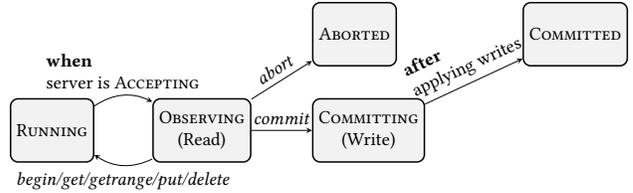
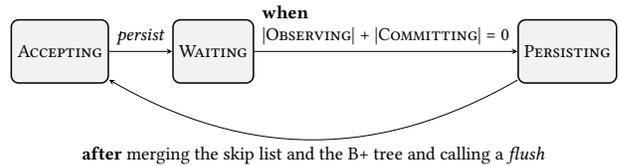
\begin{figure}[t]
  \centering
  \subfigure[Client transition diagram.]{
    \scalebox{.70}{\begin{tikzpicture} 
  \tikzset{
    ->, 
    >=stealth, 
    every state/.style={thick, fill=gray!10, minimum size=1.2cm, rectangle, rounded corners}, 
    initial text=$ $, 
  }

  \node[state] (running) {\strunning};
  \node[state, right=1.10cm of running, align=center] (observing) {\stobserving\\(Read)};
  \node[state, above right=6mm and 1.15cm of observing] (aborted) {\staborted};
  \node[state, right=1.15cm of observing, align=center] (committing) {\stcommitting\\(Write)};
  \node[state, above right=6mm and 1.85cm of committing] (committed) {\stcommitted};

  \draw (running) edge[bend left] node[above, align=left] {{\bf when}\\server is \staccepting} (observing);
  \draw (observing) edge[bend left] node[below]{\emph{begin/get/getrange/put/delete}} (running);
  \draw (observing) edge[sloped] node[above] {\emph{abort}} (aborted);
  \draw (observing) edge node[above] {\emph{commit}} (committing);
  \draw (committing) edge[sloped] node[above, align=left] {{\bf after}\\applying writes} (committed);
\end{tikzpicture}}
    \label{fig:sync-client}
  }
  \subfigure[Server transition diagram.]{
    \scalebox{.70}{\begin{tikzpicture} 
  \tikzset{
    ->, 
    >=stealth, 
    every state/.style={thick, fill=gray!10, minimum size=1.2cm, rectangle, rounded corners}, 
    initial text=$ $, 
  }

  \node[state] (accepting) {\staccepting};
  \node[state, right=1.2cm of accepting] (waiting) {\stwaiting};
  \node[state, right=5cm of waiting] (persisting) {\stpersisting};

  \draw (accepting) edge node[above] {\emph{persist}} (waiting);
  \draw (waiting) edge node[above, align=left]{ {\bf when}\\|{\stobserving}| + |{\stcommitting}| = 0} (persisting);
  \draw (persisting) edge[bend left, bend angle=5] node[below=0ex]{{\bf after} merging the skip list and the B+ tree and calling a \emph{flush}} (accepting);
\end{tikzpicture}}
    \label{fig:sync-server}
  }
  \caption{\small
Synchronizing clients and the database server.
  }
  \label{fig:sync-clients-server}
\end{figure}

\paragraph{\bf Creating consistent snapshot.}
The semantics of \syspersist gives ``consistent snapshot'' a precise definition: It is the state where committed transactions have applied all their writes, in the order governed by SS2PL (\autoref{subsec:s2pl}) to ensure serializability, and uncommitted transactions have not applied any of its writes.
Next we describe how \sys creates a consistent snapshot efficiently.

As each transaction can run for an arbitrarily long time, it would be unreasonable to wait for all active transactions to commit or abort before creating a snapshot.
Thus, \sys uses an efficient protocol to orchestrate the transactions and the database server, as depicted in the form of state transition diagrams in \autoref{fig:sync-clients-server}.

We first explain the transition diagram of a client, as shown in \autoref{fig:sync-client}.
Initially, all clients are in the \strunning state.
A client can perform database operations only in the  \stobserving state. 
So in order to perform an operation, the client has to first \emph{enter} the server by transitioning its state from \strunning to \stobserving. 
This is possible only when the server is in the \staccepting state.
An \emph{abort} operation changes the client state from \stobserving to \staborted.
A \emph{commit} operation changes the client state from \stobserving to \stcommitting.
To prevent uncommitted transactions from updating the server, each planned modification to the server has to first go to the per-transaction local write set, and is applied to the server during the client's \stcommitting phase.
In the \stcommitting state, the client updates the server with its local write set and move to \stcommitted only when all the entries in the local write set are processed.
We say that a client \emph{leaves} the server when it moves to \strunning, \staborted, or \stcommitted.

Next we explain the transition diagram of the server, as shown in \autoref{fig:sync-server}.
Initially, the server is in the \staccepting state. 
On receiving a \emph{persist}, the server first moves to the \stwaiting state and blocks all clients from entering. 
When the server is in the \stwaiting state, some clients might still in \stobserving or \stcommitting states, that is, still in the server.
The server will wait until all clients have left and then move to the \stpersisting state to create a snapshot.

The protocol ensures the property that when the server is \stpersisting, no client is \stobserving or \stcommitting.
Notice that at the moment when the server just moves to the \stpersisting state, all clients are either \strunning, \staborted, or \stcommitted.
The latter two states have no chance to transition to the \stobserving and \stcommitting states.
When the server is in the \stpersisting state, the client cannot move from \strunning to either \stobserving or \stcommitting.
Hence the property is guaranteed.
This property further ensures that only committed transactions apply their writes to the server.
For uncommitted ones, their writes only stay in their local write set.



One important assumption of the protocol is that checking transition guard conditions (e.g., checking if the state is \staccepting) and transitioning to the next state (e.g., to the \stobserving state) are done \emph{atomically}.
Without this assumption, we can easily create an example to break the protocol's property.
Assume we have only one client, and we always have one server.
Initially, the client is in the \strunning state and the server is in the \staccepting state.
The client wants to enter the server so it checks the server's state, which is now \staccepting.
Before the client moves to \stobserving, the server has moved to the \stwaiting state, checks the guard condition ``no client is \stobserving or \stcommitting'', and immediately moves to \stpersisting.
The client moves to \stobserving only after the server moves to \stpersisting.
Now the property ``when the server is \stpersisting, no client is in \stobserving or \stcommitting'' is broken.

\newcommand{\cnt}{{\ttfamily n\PYZus{}accessing}\xspace}
\newcommand{\acc}{{\ttfamily accepting}\xspace}

Thus, we develop a mechanism to ensure checking transition guards and transitioning to the next states are done atomically.
The implementation of the synchronization protocol is shown in \autoref{fig:txn-server-protocol-impl}.

\begin{figure}[h]
  \begin{minipage}{.4\columnwidth}
    \input{code/server.c.tex}
  \end{minipage}
  \hfill
  \begin{minipage}{.4\columnwidth}
    \input{code/client.c.tex}
  \end{minipage}
  \caption{Client-server protocol implementation.}
  \label{fig:txn-server-protocol-impl}
\end{figure}

When a client requests to enter the server, it increases \cnt before it actually enters the \stobserving state. 
Then it checks whether the server is in \staccepting by reading the variable \acc, and enters \stobserving if \acc is indeed true.
The transition from \stobserving to \stcommitting does not change the value of \cnt, as it represents the total number of clients in both states.
A client decreases \cnt by one when leaving the server.
Similarly, when the server attempts to enter the \stpersisting state, it first sets \acc to false, which denotes that it is already in the \stwaiting state.
Then it waits until \cnt becomes zero and then moves to \stpersisting in order to do the actual task of \syspersist.
After finishing the task, the server sets \acc to true to indicate that it has transitioned back to \staccepting.
We use memory fences to ensure the reads and writes to \cnt and \acc follow the program order.
The mechanism guarantees that (1) the server is in \staccepting at the time when client transitions to \stobserving, and (2) |{\stobserving}| + |{\stcommitting}| = 0 at the time when the server transitions to \stpersisting. 





\paragraph{\bf Discussion.}
It is known that ARIES-style WAL often performs better than shadow paging in systems where concurrent write transactions are supported~\cite{gray81:systemr-recovery, mohan92:aries}.
The reason is that with shadow paging, when a transaction is ready to write back the dirty pages at commit time, it must ensure that other transactions have not partially modified those pages, otherwise inconsistency may arise on a crash.
Consequently, transactions have to frequently synchronize with each other.
With weak durability, however, the need to urgently synchronize with other transactions no longer exists: it is only when users ask for durability that synchronization is required.

On the other hand, the main benefit of our specific shadow technique is \emph{simplicity}, in that it has a simple specification, and does not depend on layers other than the file system.
This means that we can separately test or even formally verify~\cite{chang20:scftl} the implementation against a simple specification.
In contrast, WAL often depends on other layers like buffers, or assumes a certain locking protocol, making the already difficult task of testing or verifying crash safety even harder.

\subsection{Latch-free two-level index structure}
\label{subsec:index}

\begin{figure*}[t]
  \centering
  \subfigure[Input B+-tree and sorted list.]{
    \centering
    \scalebox{.67}{\begin{tikzpicture}
  \tikzstyle{slnode} = [draw=black, fill=gray!10, minimum size=5mm, rectangle, inner sep=0pt]
  \tikzset{
    pics/sl/.style n args={3}{
      code = {
        \node[slnode, left=4mm of #1] (#2) {#3};
        \draw[->, >=stealth] (#1) -- (#2);
      }
    }
  }
  \tikzstyle{treenode} = [
    minimum width=7mm, minimum height=5mm,
    rectangle, inner sep=0pt]
  \tikzset{
    opaque/.style={opacity=.4},
    pics/btreeopa/.style n args={4}{
      code = {
        \node[treenode, opaque] (#1y) {#3};
        \node[treenode, opaque, left=0pt of #1y] (#1x) {#2};
        \node[treenode, opaque, right=0pt of #1y] (#1z) {#4};
        \begin{scope}[on background layer]
          \fill[blue!10, draw=black, opaque] (#1x.north west) rectangle (#1z.south east);
        \end{scope}
      }
    },
    pics/btree/.style n args={4}{
      code = {
        \node[treenode] (#1y) {#3};
        \node[treenode, left=0pt of #1y] (#1x) {#2};
        \node[treenode, right=0pt of #1y] (#1z) {#4};
        \begin{scope}[on background layer]
          \fill[blue!10, draw=black] (#1x.north west) rectangle (#1z.south east);
        \end{scope}
      }
    },
  }

  \small
  \draw pic {btree={root}{24}{max}{}};
  \draw pic[below left=3mm and 1cm of rooty.south] {btree={b1}{24}{}{}};
  \draw pic[below right=3mm and 1cm of rooty.south] {btree={b2}{35}{max}{}};
  \draw pic[below left=3mm and 1cm of b1y.south] {btree={b3}{17}{24}{}};
  \draw pic[below right=3mm and 1cm of b1y.south] {btree={b4}{35}{}{}};
  \draw pic[below right=3mm and 1cm of b2y.south] {btree={b5}{37}{50}{}};
  \draw[->, >=stealth] (rootx.south) -- (b1y.north);
  \draw[->, >=stealth] (rooty.south) -- (b2y.north);
  \draw[->, >=stealth] (b1x.south) -- (b3y.north);
  \draw[->, >=stealth] (b2x.south) -- (b4y.north);
  \draw[->, >=stealth] (b2y.south) -- (b5y.north);
  \draw[->, >=stealth] (b4x) -- (b3z);
  \draw[->, >=stealth] (b5x) -- (b4z);
  \node[below=4mm of b5x, slnode] (n0) {40};
  \draw pic {sl={n0}{n1}{36}};
  \draw pic {sl={n1}{n2}{31}};
  \draw pic {sl={n2}{n3}{18}};
  \draw pic {sl={n3}{n4}{14}};
  \draw pic {sl={n4}{n5}{8}};
  \node[right=4mm of n0] (h0) {};
  \draw[->, >=stealth] (h0) -- (n0) {};

\end{tikzpicture}}
  }
  \subfigure[Partitioning (leaf).]{
    \centering
    \scalebox{.67}{\begin{tikzpicture}
  \tikzstyle{slnode} = [draw=black, fill=gray!10, minimum size=5mm, rectangle, inner sep=0pt]
  \tikzset{
    pics/sl/.style n args={4}{
      code = {
        \node[slnode, #4=3mm of #2] (#1) {#3};
        \ifthenelse
          {\equal{#4}{right}}
          {\draw[->, >=stealth] (#2) -- (#1);}
          {\draw[->, >=stealth] (#1) -- (#2);};
      }
    }
  }
  \tikzstyle{treenode} = [
    minimum width=7mm, minimum height=5mm,
    rectangle, inner sep=0pt]
  \tikzset{
    opaque/.style={opacity=.4},
    pics/btreeopa/.style n args={4}{
      code = {
        \node[treenode, opaque] (#1y) {#3};
        \node[treenode, opaque, left=0pt of #1y] (#1x) {#2};
        \node[treenode, opaque, right=0pt of #1y] (#1z) {#4};
        \begin{scope}[on background layer]
          \fill[blue!10, draw=black, opaque] (#1x.north west) rectangle (#1z.south east);
        \end{scope}
      }
    },
    pics/btree/.style n args={4}{
      code = {
        \node[treenode] (#1y) {#3};
        \node[treenode, left=0pt of #1y] (#1x) {#2};
        \node[treenode, right=0pt of #1y] (#1z) {#4};
        \begin{scope}[on background layer]
          \fill[blue!10, draw=black] (#1x.north west) rectangle (#1z.south east);
        \end{scope}
      }
    },
  }

  \small
  \draw pic {btreeopa={root}{24}{max}{}};
  \draw pic[below left=3mm and 1cm of rooty.south] {btreeopa={b1}{24}{}{}};
  \draw pic[below right=3mm and 1cm of rooty.south] {btreeopa={b2}{35}{max}{}};
  \draw pic[below left=3mm and 1cm of b1y.south] {btree={b3}{17}{24}{}};
  \draw pic[below right=3mm and 1cm of b1y.south] {btree={b4}{35}{}{}};
  \draw pic[below right=3mm and 1cm of b2y.south] {btree={b5}{37}{50}{}};
  \draw[->, >=stealth, opaque] (rootx.south) -- (b1y.north);
  \draw[->, >=stealth, opaque] (rooty.south) -- (b2y.north);
  \draw[->, >=stealth, opaque] (b1x.south) -- (b3y.north);
  \draw[->, >=stealth, opaque] (b2x.south) -- (b4y.north);
  \draw[->, >=stealth, opaque] (b2y.south) -- (b5y.north);
  \draw[->, >=stealth] (b4x) -- (b3z);
  \draw[->, >=stealth] (b5x) -- (b4z);
  \node[below = 4mm of b3x, slnode] (n0) {8};
  \node[below = 4mm of b3y, slnode] (n1) {14};
  \node[below = 4mm of b3z, slnode] (n2) {18};
  \node[below = 4mm of b4y, slnode] (n3) {31};
  \node[below = 4mm of b5x.south east, slnode] (n4) {36};
  \node[below = 4mm of b5z.south west, slnode] (n5) {40};
  \node[fit=(n0)(n1)(n2)(b3x)(b3z), draw=black, dashed] {};
  \node[fit=(n3)(b4x)(b4z), draw=black, dashed] {};
  \node[fit=(n4)(n5)(b5x)(b5z), draw=black, dashed] {};

\end{tikzpicture}}
  }
  \subfigure[Coalescing (leaf).]{
    \centering
    \scalebox{.67}{\begin{tikzpicture}
  \tikzstyle{slnode} = [draw=black, fill=gray!10, minimum size=5mm, rectangle, inner sep=0pt]
  \tikzset{
    pics/sl/.style n args={4}{
      code = {
        \node[slnode, #4=3mm of #2] (#1) {#3};
        \ifthenelse
          {\equal{#4}{right}}
          {\draw[->, >=stealth] (#2) -- (#1);}
          {\draw[->, >=stealth] (#1) -- (#2);};
      }
    }
  }
  \tikzstyle{treenode} = [
    minimum width=7mm, minimum height=5mm,
    rectangle, inner sep=0pt]
  \tikzstyle{gptrnode} = [
    draw=black, fill=green!10, minimum size=5mm, rectangle
  ]
  \tikzset{
    opaque/.style={opacity=.4},
    pics/btreeopa/.style n args={4}{
      code = {
        \node[treenode, opaque] (#1y) {#3};
        \node[treenode, opaque, left=0pt of #1y] (#1x) {#2};
        \node[treenode, opaque, right=0pt of #1y] (#1z) {#4};
        \begin{scope}[on background layer]
          \fill[blue!10, draw=black, opaque] (#1x.north west) rectangle (#1z.south east);
        \end{scope}
      }
    },
    pics/btree/.style n args={4}{
      code = {
        \node[treenode] (#1y) {#3};
        \node[treenode, left=0pt of #1y] (#1x) {#2};
        \node[treenode, right=0pt of #1y] (#1z) {#4};
        \begin{scope}[on background layer]
          \fill[blue!10, draw=black] (#1x.north west) rectangle (#1z.south east);
        \end{scope}
      }
    },
  }

  \small
  \draw pic {btreeopa={root}{24}{max}{}};
  \draw pic[below left=3mm and 1cm of rooty.south] {btreeopa={b1}{24}{}{}};
  \draw pic[below right=3mm and 1cm of rooty.south] {btreeopa={b2}{35}{max}{}};
  \draw pic[below left=3mm and 1cm of b1y.south] {btree={b3}{18}{24}{}};
  \draw pic[below right=3mm and 1cm of b1y.south] {btree={b4}{31}{35}{}};
  \draw pic[below right=3mm and 1cm of b2y.south] {btree={b5}{40}{50}{}};
  \draw[->, >=stealth, opaque] (rootx.south) -- (b1y.north);
  \draw[->, >=stealth, opaque] (rooty.south) -- (b2y.north);
  \draw[->, >=stealth, opaque] (b1x.south) -- (b3y.north);
  \draw[->, >=stealth, opaque] (b2x.south) -- (b4y.north);
  \draw[->, >=stealth, opaque] (b2y.south) -- (b5y.north);
  \draw[->, >=stealth] (b4x) -- (b3z);
  \draw pic[below=2mm of b3y] {btree={b6}{14}{17}{}};
  \draw pic[below=2mm of b6y] {btree={b7}{8}{}{}};
  \draw (b3z.east) edge[bend left, ->, >=stealth] (b6z.east);
  \draw (b6z.east) edge[bend left, ->, >=stealth] (b7z.east);
  \draw pic[below=2mm of b5y] {btree={b8}{36}{37}{}};
  \draw (b5z.east) edge[bend left, ->, >=stealth] (b8z.east);
  \draw (b8x.west) edge[out=180, in=0, ->, >=stealth, looseness=2] (b4z.east);
  \node[left=3mm of b6x, gptrnode] (g0) {17};
  \node[left=3mm of b7x, gptrnode] (g1) {8};
  \node[right=3mm of b8z, gptrnode] (g2) {37};
  \draw[->, >=stealth] (g0) -- (b6x);
  \draw[->, >=stealth] (g1) -- (b7x);
  \draw[->, >=stealth] (g2) -- (b8z);

\end{tikzpicture}}
  }
  \subfigure[Collecting (leaf).]{
    \centering
    \scalebox{.67}{\begin{tikzpicture}
  \tikzstyle{slnode} = [draw=black, fill=gray!10, minimum size=5mm, rectangle, inner sep=0pt]
  \tikzset{
    pics/sl/.style n args={4}{
      code = {
        \node[slnode, #4=3mm of #2] (#1) {#3};
        \ifthenelse
          {\equal{#4}{right}}
          {\draw[->, >=stealth] (#2) -- (#1);}
          {\draw[->, >=stealth] (#1) -- (#2);};
      }
    }
  }
  \tikzstyle{treenode} = [
    minimum width=7mm, minimum height=5mm,
    rectangle, inner sep=0pt]
  \tikzstyle{gptrnode} = [
    draw=black, fill=green!10, minimum size=5mm, rectangle
  ]
  \tikzset{
    opaque/.style={opacity=.4},
    pics/btreeopa/.style n args={4}{
      code = {
        \node[treenode, opaque] (#1y) {#3};
        \node[treenode, opaque, left=0pt of #1y] (#1x) {#2};
        \node[treenode, opaque, right=0pt of #1y] (#1z) {#4};
        \begin{scope}[on background layer]
          \fill[blue!10, draw=black, opaque] (#1x.north west) rectangle (#1z.south east);
        \end{scope}
      }
    },
    pics/btree/.style n args={4}{
      code = {
        \node[treenode] (#1y) {#3};
        \node[treenode, left=0pt of #1y] (#1x) {#2};
        \node[treenode, right=0pt of #1y] (#1z) {#4};
        \begin{scope}[on background layer]
          \fill[blue!10, draw=black] (#1x.north west) rectangle (#1z.south east);
        \end{scope}
      }
    },
  }

  \small
  \draw pic {btreeopa={root}{24}{max}{}};
  \draw pic[below left=3mm and 1cm of rooty.south] {btree={b1}{24}{}{}};
  \draw pic[below right=3mm and 1cm of rooty.south] {btree={b2}{35}{max}{}};
  \draw[->, >=stealth, opaque] (rootx.south) -- (b1y.north);
  \draw[->, >=stealth, opaque] (rooty.south) -- (b2y.north);
  \node[below=4mm of $(b1z.south east)!0.5!(b2x.south west)$, gptrnode] (g1) {17};
  \node[right=4mm of g1, gptrnode] (g0) {37};
  \node[left=4mm of g1, gptrnode] (g2) {8};
  \draw[->, >=stealth] (g0) -- (g1);
  \draw[->, >=stealth] (g1) -- (g2);
  \draw[->, >=stealth] ($(g0.east) + (4mm, 0)$) -- (g0);

\end{tikzpicture}}
  }
  \subfigure[Partitioning (level 1).]{
    \centering
    \scalebox{.67}{\begin{tikzpicture}
  \tikzstyle{slnode} = [draw=black, fill=gray!10, minimum size=5mm, rectangle, inner sep=0pt]
  \tikzset{
    pics/sl/.style n args={4}{
      code = {
        \node[slnode, #4=3mm of #2] (#1) {#3};
        \ifthenelse
          {\equal{#4}{right}}
          {\draw[->, >=stealth] (#2) -- (#1);}
          {\draw[->, >=stealth] (#1) -- (#2);};
      }
    }
  }
  \tikzstyle{treenode} = [
    minimum width=7mm, minimum height=5mm,
    rectangle, inner sep=0pt]
  \tikzstyle{gptrnode} = [
    draw=black, fill=green!10, minimum size=5mm, rectangle
  ]
  \tikzset{
    opaque/.style={opacity=.4},
    pics/btreeopa/.style n args={4}{
      code = {
        \node[treenode, opaque] (#1y) {#3};
        \node[treenode, opaque, left=0pt of #1y] (#1x) {#2};
        \node[treenode, opaque, right=0pt of #1y] (#1z) {#4};
        \begin{scope}[on background layer]
          \fill[blue!10, draw=black, opaque] (#1x.north west) rectangle (#1z.south east);
        \end{scope}
      }
    },
    pics/btree/.style n args={4}{
      code = {
        \node[treenode] (#1y) {#3};
        \node[treenode, left=0pt of #1y] (#1x) {#2};
        \node[treenode, right=0pt of #1y] (#1z) {#4};
        \begin{scope}[on background layer]
          \fill[blue!10, draw=black] (#1x.north west) rectangle (#1z.south east);
        \end{scope}
      }
    },
  }

  \small
  \draw pic {btreeopa={root}{24}{max}{}};
  \draw pic[below left=3mm and 1cm of rooty.south] {btree={b1}{24}{}{}};
  \draw pic[below right=3mm and 1cm of rooty.south] {btree={b2}{35}{max}{}};
  \draw[->, >=stealth, opaque] (rootx.south) -- (b1y.north);
  \draw[->, >=stealth, opaque] (rooty.south) -- (b2y.north);
  \node[below=4mm of b2y, gptrnode] (g0) {37};
  \node[below=4mm of b1z.south west, gptrnode] (g1) {17};
  \node[below=4mm of b1x.south east, gptrnode] (g2) {8};
  \node[fit=(g0)(b2x)(b2z), draw=black, dashed] {};
  \node[fit=(g2)(g1)(b1x)(b1z), draw=black, dashed] {};

\end{tikzpicture}}
  }
  \subfigure[Coalescing (level 1).]{
    \centering
    \scalebox{.67}{\begin{tikzpicture}
  \tikzstyle{slnode} = [draw=black, fill=gray!10, minimum size=5mm, rectangle, inner sep=0pt]
  \tikzset{
    pics/sl/.style n args={4}{
      code = {
        \node[slnode, #4=3mm of #2] (#1) {#3};
        \ifthenelse
          {\equal{#4}{right}}
          {\draw[->, >=stealth] (#2) -- (#1);}
          {\draw[->, >=stealth] (#1) -- (#2);};
      }
    }
  }
  \tikzstyle{treenode} = [
    minimum width=7mm, minimum height=5mm,
    rectangle, inner sep=0pt]
  \tikzstyle{gptrnode} = [
    draw=black, fill=brown!10, minimum size=5mm, rectangle
  ]
  \tikzset{
    opaque/.style={opacity=.4},
    pics/btreeopa/.style n args={4}{
      code = {
        \node[treenode, opaque] (#1y) {#3};
        \node[treenode, opaque, left=0pt of #1y] (#1x) {#2};
        \node[treenode, opaque, right=0pt of #1y] (#1z) {#4};
        \begin{scope}[on background layer]
          \fill[blue!10, draw=black, opaque] (#1x.north west) rectangle (#1z.south east);
        \end{scope}
      }
    },
    pics/btree/.style n args={4}{
      code = {
        \node[treenode] (#1y) {#3};
        \node[treenode, left=0pt of #1y] (#1x) {#2};
        \node[treenode, right=0pt of #1y] (#1z) {#4};
        \begin{scope}[on background layer]
          \fill[blue!10, draw=black] (#1x.north west) rectangle (#1z.south east);
        \end{scope}
      }
    },
  }

  \small
  \draw pic {btreeopa={root}{24}{max}{}};
  \draw pic[below left=3mm and 1cm of rooty.south] {btree={b1}{17}{24}{}};
  \draw pic[below right=3mm and 1cm of rooty.south] {btree={b2}{37}{max}{}};
  \draw[->, >=stealth, opaque] (rootx.south) -- (b1y.north);
  \draw[->, >=stealth, opaque] (rooty.south) -- (b2y.north);
  \draw pic[below=2mm of b1y] {btree={b3}{8}{}{}};
  \draw pic[below=2mm of b2y] {btree={b4}{35}{}{}};
  \node[left=3mm of b3x, gptrnode] (g0) {8};
  \draw[->, >=stealth] (g0) -- (b3x);
  \node[right=3mm of b4z, gptrnode] (g1) {35};
  \draw[->, >=stealth] (g1) -- (b4z);

\end{tikzpicture}}
  }
  \subfigure[Old root.]{
    \centering
    \scalebox{.67}{\begin{tikzpicture}
  \tikzstyle{slnode} = [draw=black, fill=gray!10, minimum size=5mm, rectangle, inner sep=0pt]
  \tikzset{
    pics/sl/.style n args={4}{
      code = {
        \node[slnode, #4=3mm of #2] (#1) {#3};
        \ifthenelse
          {\equal{#4}{right}}
          {\draw[->, >=stealth] (#2) -- (#1);}
          {\draw[->, >=stealth] (#1) -- (#2);};
      }
    }
  }
  \tikzstyle{treenode} = [
    minimum width=7mm, minimum height=5mm,
    rectangle, inner sep=0pt]
  \tikzstyle{gptrnode} = [
    draw=black, fill=brown!10, minimum size=5mm, rectangle
  ]
  \tikzset{
    pics/btree/.style n args={4}{
      code = {
        \node[treenode] (#1y) {#3};
        \node[treenode, left=0pt of #1y] (#1x) {#2};
        \node[treenode, right=0pt of #1y] (#1z) {#4};
        \begin{scope}[on background layer]
          \fill[blue!10, draw=black] (#1x.north west) rectangle (#1z.south east);
        \end{scope}
      }
    },
  }

  \small
  \draw pic {btree={root}{24}{max}{}};
  \node[below=4mm of rootx.south east, gptrnode] (g0) {8};
  \node[below=4mm of rootz.south west, gptrnode] (g1) {35};
  \node[fit=(g0)(g1)(rootx)(rootz), draw=black, dashed, rectangle] {};

\end{tikzpicture}}
  }
  \subfigure[New root.]{
    \centering
    \scalebox{.67}{\begin{tikzpicture}
  \tikzstyle{slnode} = [draw=black, fill=gray!10, minimum size=5mm, rectangle, inner sep=0pt]
  \tikzset{
    pics/sl/.style n args={4}{
      code = {
        \node[slnode, #4=3mm of #2] (#1) {#3};
        \ifthenelse
          {\equal{#4}{right}}
          {\draw[->, >=stealth] (#2) -- (#1);}
          {\draw[->, >=stealth] (#1) -- (#2);};
      }
    }
  }
  \tikzstyle{treenode} = [
    minimum width=7mm, minimum height=5mm,
    rectangle, inner sep=0pt]
  \tikzstyle{gptrnode} = [
    draw=black, fill=brown!10, minimum size=5mm, rectangle
  ]
  \tikzset{
    pics/btree/.style n args={4}{
      code = {
        \node[treenode] (#1y) {#3};
        \node[treenode, left=0pt of #1y] (#1x) {#2};
        \node[treenode, right=0pt of #1y] (#1z) {#4};
        \begin{scope}[on background layer]
          \fill[blue!10, draw=black] (#1x.north west) rectangle (#1z.south east);
        \end{scope}
      }
    },
  }

  \small
  \draw pic {btree={root}{24}{max}{}};
  \draw pic[below left=3mm and 1cm of rooty.south] {btree={b1}{8}{24}{}};
  \draw pic[below right=3mm and 1cm of rooty.south] {btree={b2}{35}{max}{}};
  \draw[->, >=stealth] (rootx.south) -- (b1y.north);
  \draw[->, >=stealth] (rooty.south) -- (b2y.north);

\end{tikzpicture}}
  }
  \subfigure[Output B+-tree.]{
    \centering
    \scalebox{.67}{\begin{tikzpicture}
  \tikzstyle{slnode} = [draw=black, fill=gray!10, minimum size=5mm, rectangle, inner sep=0pt]
  \tikzset{
    pics/sl/.style n args={4}{
      code = {
        \node[slnode, #4=3mm of #2] (#1) {#3};
        \ifthenelse
          {\equal{#4}{right}}
          {\draw[->, >=stealth] (#2) -- (#1);}
          {\draw[->, >=stealth] (#1) -- (#2);};
      }
    }
  }
  \tikzstyle{treenode} = [
    minimum width=7mm, minimum height=5mm,
    rectangle, inner sep=0pt]
  \tikzstyle{gptrnode} = [
    draw=black, fill=green!10, minimum size=5mm, rectangle
  ]
  \tikzset{
    opaque/.style={opacity=.4},
    pics/btreeopa/.style n args={4}{
      code = {
        \node[treenode, opaque] (#1y) {#3};
        \node[treenode, opaque, left=0pt of #1y] (#1x) {#2};
        \node[treenode, opaque, right=0pt of #1y] (#1z) {#4};
        \begin{scope}[on background layer]
          \fill[blue!10, draw=black, opaque] (#1x.north west) rectangle (#1z.south east);
        \end{scope}
      }
    },
    pics/btree/.style n args={4}{
      code = {
        \node[treenode] (#1y) {#3};
        \node[treenode, left=0pt of #1y] (#1x) {#2};
        \node[treenode, right=0pt of #1y] (#1z) {#4};
        \begin{scope}[on background layer]
          \fill[blue!10, draw=black] (#1x.north west) rectangle (#1z.south east);
        \end{scope}
      }
    },
  }

  \small
  \draw pic {btree={root}{24}{max}{}};
  \draw pic[below left=3mm and 1cm of rooty.south] {btree={b1}{8}{24}{}};
  \draw pic[below right=3mm and 1cm of rooty.south] {btree={b2}{35}{max}{}};
  \draw pic[below left=3mm and 23.5mm of b1y.south] {btree={b3}{8}{}{}};
  \draw pic[below=3mm of b1y.south] {btree={b4}{17}{24}{}};
  \draw pic[below=3mm of b2y.south] {btree={b5}{35}{}{}};
  \draw pic[below right=3mm and 23.5mm of b2y.south] {btree={b6}{37}{max}{}};
  \draw pic[below left=3mm and 23.5mm of b3y.south] {btree={b7}{8}{}{}};
  \draw pic[below=3mm of b3y.south] {btree={b8}{14}{17}{}};
  \draw pic[below=3mm of b4y.south] {btree={b9}{18}{24}{}};
  \draw pic[below=3mm of b5y.south] {btree={b10}{31}{35}{}};
  \draw pic[below=3mm of b6y.south] {btree={b11}{36}{37}{}};
  \draw pic[below right=3mm and 23.5mm of b6y.south] {btree={b12}{40}{50}{}};
  \draw[->, >=stealth] (rootx.south) -- (b1y.north);
  \draw[->, >=stealth] (rooty.south) -- (b2y.north);
  \draw[->, >=stealth] (b1x.south) -- (b3y.north);
  \draw[->, >=stealth] (b1y.south) -- (b4y.north);
  \draw[->, >=stealth] (b2x.south) -- (b5y.north);
  \draw[->, >=stealth] (b2y.south) -- (b6y.north);
  \draw[->, >=stealth] (b3x.south) -- (b7y.north);
  \draw[->, >=stealth] (b4x.south) -- (b8y.north);
  \draw[->, >=stealth] (b4y.south) -- (b9y.north);
  \draw[->, >=stealth] (b5x.south) -- (b10y.north);
  \draw[->, >=stealth] (b6x.south) -- (b11y.north);
  \draw[->, >=stealth] (b6y.south) -- (b12y.north);
  \draw[->, >=stealth] (b12x) -- (b11z);
  \draw[->, >=stealth] (b11x) -- (b10z);
  \draw[->, >=stealth] (b10x) -- (b9z);
  \draw[->, >=stealth] (b9x) -- (b8z);
  \draw[->, >=stealth] (b8x) -- (b7z);

\end{tikzpicture}}
  }
  \caption{\small
Merging the in-memory skip list into the on-disk B+-tree (each number shown in the figure represents a record, but we only show its key and omit its value).
(a) The merging procedure takes a B+-tree and a sorted list as input.
(b) Each record is assigned to the tree node whose range covers the key of the record; for instance, as the middle leaf node should contain keys within $(24, 35]$, the record with key $31$ is assigned to this node.
(c) After the assignment, records are coalesced into each tree node.
For nodes that still have space (\eg the middle one), simple insertions suffice;
otherwise, a merge sort is performed on the records to be inserted and records in the tree node.
During the merge sort, new tree nodes are allocated and internal pointers (colored in green) pointing to the new nodes are generated.
(d) The internal pointers are then collected to form the input of level 1.
(e and f) The same procedure is then applied to level 1.
(g) We have reached the root, but the root cannot accommodate the internal pointers (colored in brown), (h) so a new root is created.
(i) shows the output B+-tree.
  }
  \label{fig:merge}
\end{figure*}
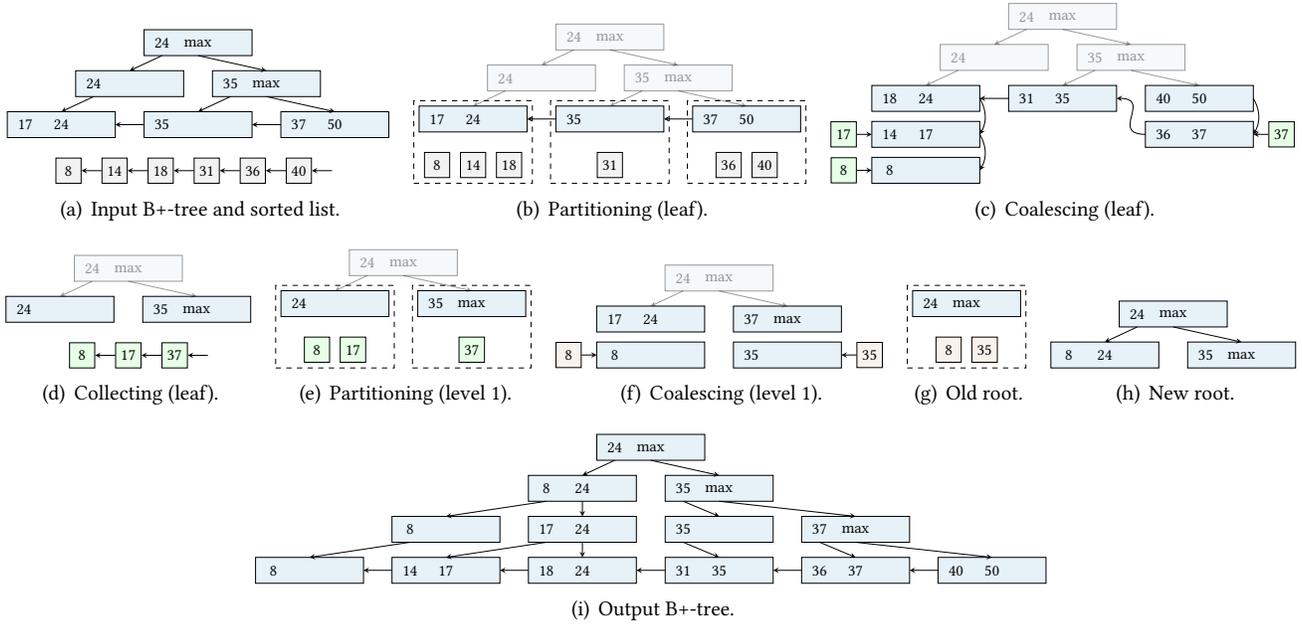

Weak durability offers a great opportunity for \emph{batch processing}---operations within two consecutive {\syspersist}s naturally form a batch.
Moreover, weak durability frees us from worrying about the durability of operations generated in the last batch.
\sys embodies these ideas in its index design by employing a latch-free two-level index structure that scales well across a wide range of workloads.

As shown in \autoref{fig:design-overview}, the index consists of an in-memory latch-free concurrent skip list~\cite{pugh90:skiplist, fraser04:practical} and an on-disk B+-tree.
The interesting part of the two-level design is that it only supports merging a batch of records, rather than inserting a single record, into the B+-tree.
Single-record insertions will go to the skip list.
As insertions of records are absorbed by the skip list, and deletions, like many other systems~\cite{raju17:pebblesdb, arulraj15:nvm-recovery}, are implemented as a special form of updates that set the records to a tombstone value, the B+-tree structure is guaranteed to remain the same within the same batch.
This \emph{structural invariance property} is appealing for two primary reasons, as described below.

First, the index does not require latches to protect the tree structure; along with the fact that the implementation of a concurrent skip list is simpler and more scalable than that of a concurrent B+-tree~\cite{pugh90:skiplist}, this combination leads to better scalability.
Second, records are guaranteed to remain at the same location within the same batch, which fits well with our use of the local write set (\autoref{subsec:shadow}).
With this guarantee, a transaction can now simply store the location of records they wish to update in its write set, and at commit time, apply the writes directly to the stored location if the batch has not changed since the time the transaction retrieves the record.
This removes the need to relocate records at commit time in most cases.

\paragraph{\bf Merging skip list into B+-tree.}
On receiving a \syspersist, \sys merges the in-memory skip list into the on-disk B+-tree for durability.
The merging algorithm targets one level at a time, starting from the leaf level and going upward to the root.
At each level, the merging proceeds in three phases: \emph{partitioning}, \emph{coalescing}, and \emph{collecting}, as illustrated in \autoref{fig:merge} with an example.
First, the input list is partitioned into multiple sublists, with each sublist containing keys to be coalesced into the same B+-tree node.
Then, for each sublist, a dedicated worker thread inserts each record in the sublist into the tree node when there is enough space; otherwise, it performs a merge sort on the tree node and the sublist.
When a new pointer is generated due to a split, it does not immediately propagate to the upper levels.
Rather, the new pointer is collected in an output list, and linked with other output lists to produce the input for the next level.

A similar (and more detailed) merging algorithm is also described in~\cite{sewall11:palm}, along with some optimizations such as balancing the inserting load and utilizing SIMD instructions to speed up sorting.

\paragraph{\bf Discussion.}
The multi-level design of \sys's index shares a similar structure to log-structured merge-trees (LSM-trees)~\cite{oneil96:lsm-tree}, and also appears in some research prototypes~\cite{stonebraker05:cstore} and commercial systems~\cite{chang06:bigtable}.
But there are some notable differences in the index design of \sys and of prior systems:
First, \sys only inserts new records into the skip list, and directly modifies records that are already present in the skip list or in the B+-tree.
In contrast, other systems direct both insertions and modifications to their write-efficient data structure.
As a result, the merging load of \sys is smaller than that of prior systems when the workload involves a few updates.
Another difference is that \sys does not need a protective mechanism to ensure that the records in the skip list are not lost upon a crash, as they are inserted \emph{after} the last \syspersist.
The permissiveness of weak durability is the key to enable this design choice.

\subsection{Strong strict two-phase locking (SS2PL)}
\label{subsec:s2pl}

To ensure serializability, \sys follows the standard SS2PL locking protocol to acquire shared and exclusive locks on records and on key gaps~\cite{mohan90:next-key}.
For completeness, we briefly describe the main idea here.

Before reading and writing a record, each transaction must obtain the shared and, respectively, exclusive record-lock on the key of the record.
Additionally, a range query also has to make sure that the key range it concerns will not be affected by future insertions.
This is done with a special form of locks called \emph{gap-locks}, which serve as ``physical surrogates for logical properties''~\cite{hellerstein07:arch-db}.
This means that by obtaining a gap-lock on key $k$ (a physical entity), the transaction essentially owns the range between $k$ and the key immediately preceding $k$ (a logical range).
For instance, if a transaction wants to retrieve records whose key lies between $3$ and $6$, and currently the database has records with key $1$, $4$, and $8$, then the transaction would have to acquire shared gap-locks on $4$ and $8$.
Similarly, a transaction has to acquire an exclusive gap-lock for the range into which it wishes to insert the record.
If the key exceeds the largest key currently existing in the database, then the transaction would have to acquire a gap-lock on a special sentinel key, which, in effect, locks the entire rightmost range.
To avoid deadlocks, \sys adopts the no-wait policy~\cite{yu14:cc-thousand}; that is, when a transaction fails to acquire a lock, it simply aborts.

\paragraph{Prefix preservation of SS2PL.}
As mentioned in \autoref{subsec:crash-consistency}, to ensure crash consistency, we can use a scheduling mechanism that possesses the prefix preservation property (in addition to serializability).
SS2PL indeed has this property, as it requires every transaction to hold its locks (both shared and exclusive) until the transaction terminates; consequently, any operation $\mathit{op}$ that depends on another operation $\mathit{op}'$ of a different transaction can only take effect after the transaction executing $\mathit{op}'$ commits.

\subsection{Top-level operations}
\label{subsec:top-op}

To put the pieces together, we go through how \sys handles each primitive.

\paragraph{\sysget.}
The transaction first acquires a shared record-lock on $k$, searches for $k$ in the local write set and then the index, and returns $v$ if $(k, v)$ is found.

\paragraph{\sysgetrange.}
The transaction first acquires a shared gap-lock on the smallest key that is greater than or equal to $k2$ in the index.
Then it retrieves all records whose key belongs to the range $[k1, k2]$ from the local write set and the index, acquires a shared gap-lock and a shared record-lock on each record, and returns the retrieved records.

\paragraph{\sysput.}
The transaction first checks whether $k$ is present in the local write set; if so, it directly updates the corresponding entry and returns.
If not, it acquires an exclusive record-lock on $k$, and search for key $k$ in the index.
If $k$ is present in the index, it stores $(k, v)$ together with the location (which can be a node address, tagged as \wsetlist, or a B+-tree record location, tagged as \wsettree) in its local write set.
Otherwise, it acquires an exclusive gap-lock on the smallest key greater than or equal to $k$, allocate a local write set entry to store $(k, v)$, and tags the location of this entry as \wsetnone.

\paragraph{\sysdelete.}
The transaction acquires an exclusive record-lock on $k$, and searchs for $k$ in the local write set and the index.
If $k$ is found, it effectively performs a \emph{put} but the update value is set to a tombstone value; in \sys, a tombstone value is any value whose length is zero.
Otherwise, the transaction does nothing.

\paragraph{\sysbegin.}
The transaction simply records the current epoch.

\paragraph{\syscommit.}
The transaction compares the current epoch with the one recorded in \sysbegin.
If they match, the transaction directly applies the write set to the skip list or the B+-tree based on the location information stored in the local write set.
If they do not match, meaning that a \syspersist has happened in the midst of \sysbegin and \syscommit, the locations are thus invalid as \syspersist has merged the skip list into the B+-tree.
In this case, the transaction has to search for key $k$ in the B+-tree for each write set entry which previously resides in the index (\ie an entry whose location is tagged as \wsetlist or \wsettree) to find out their new locations.
With the new locations, the transaction can now apply the write set to the index.
Finally, it releases all the acquired locks and pinned buffers, and resets the local write set.

\paragraph{\sysabort.}
The transaction simply releases all the acquired locks and pinned buffers, and resets the local write set.

\paragraph{\syspersist.}
The transaction merges the skip list into the B+-tree, writes back all the dirty buffers to the shadow paging layer, calls a \scflush to crash-atomically update the stable B+-tree, and finally advances the current epoch.

%
%
%



\section{Evaluation}
\label{sec:eval}

We ran experiments to answer the following questions:
\begin{denseitemize}
\item What are the trade-offs imposed by weak durability and by group commit? (\autoref{subsec:group-commit})
\item How does \sys perform compared with existing disk-based database systems? (\autoref{subsec:ycsb})
\item What are the benefits of building \sys around weak durability? (\autoref{subsec:multithreading})
\item What is the recovery time (\autoref{subsec:recovery-time}) and memory overhead (\autoref{subsec:memory-overhead}) of \sys?
\end{denseitemize}

\subsection{Experimental setup}
\label{subsec:exp-setup}

All experiments were done on a host machine with a 6-core 3.2 GHz Intel i7-8700 CPU and 16 GB of DRAM.
All database files were stored on an ext4 file system hosted by a Samsung 980 Pro 500-GB NVMe SSD (except for \autoref{subsec:group-commit} where we also use an HDD).
We use the YCSB benchmark~\cite{cooper10:ycsb} to evaluate the performance of \sys. 
The database contains 20M records, each of which consists of a 16B key and a 100B value.
Details of each workload are described below:
\paragraph{\bf Read-or-write.}
Each transaction reads or writes a record whose key is chosen randomly and uniformly from the entire database.
The read ratio $r$ controls the percentage of read operations.
\paragraph{\bf Insertion.}
Each transaction inserts a record whose key is chosen randomly and uniformly from $[0, $ 20M$)$.
For this workload, the database is initially empty.
\paragraph{\bf Range query.}
Each transaction randomly and uniformly selects a key $k$ from the entire database, and reads $n$ records starting from $k$, where $n$ is chosen randomly and uniformly from $[1, 100]$.
\paragraph{\bf Read-modify-write.}
Each transaction reads a record whose key is chosen randomly and uniformly from the entire database, and then updates the value of the record.

\subsection{Weak durability versus group commit}
\label{subsec:group-commit}

Weak durability and group commit share a similar goal: alleviating the high synchronization overhead due to the strong durability requirement.
One interesting fact about \sys's interface is that it can easily support group commit by delaying the notification that a \syscommit has returned until the next \syspersist.
This way, transactions between two consecutive {\syspersist}s naturally form a batch.
This gives us a great opportunity to compare weak durability with group commit on a similar experimental setting, to better understand the trade-off imposed by each technique.

\begin{figure}[t]
  \centering
  \includegraphics[width=\linewidth]{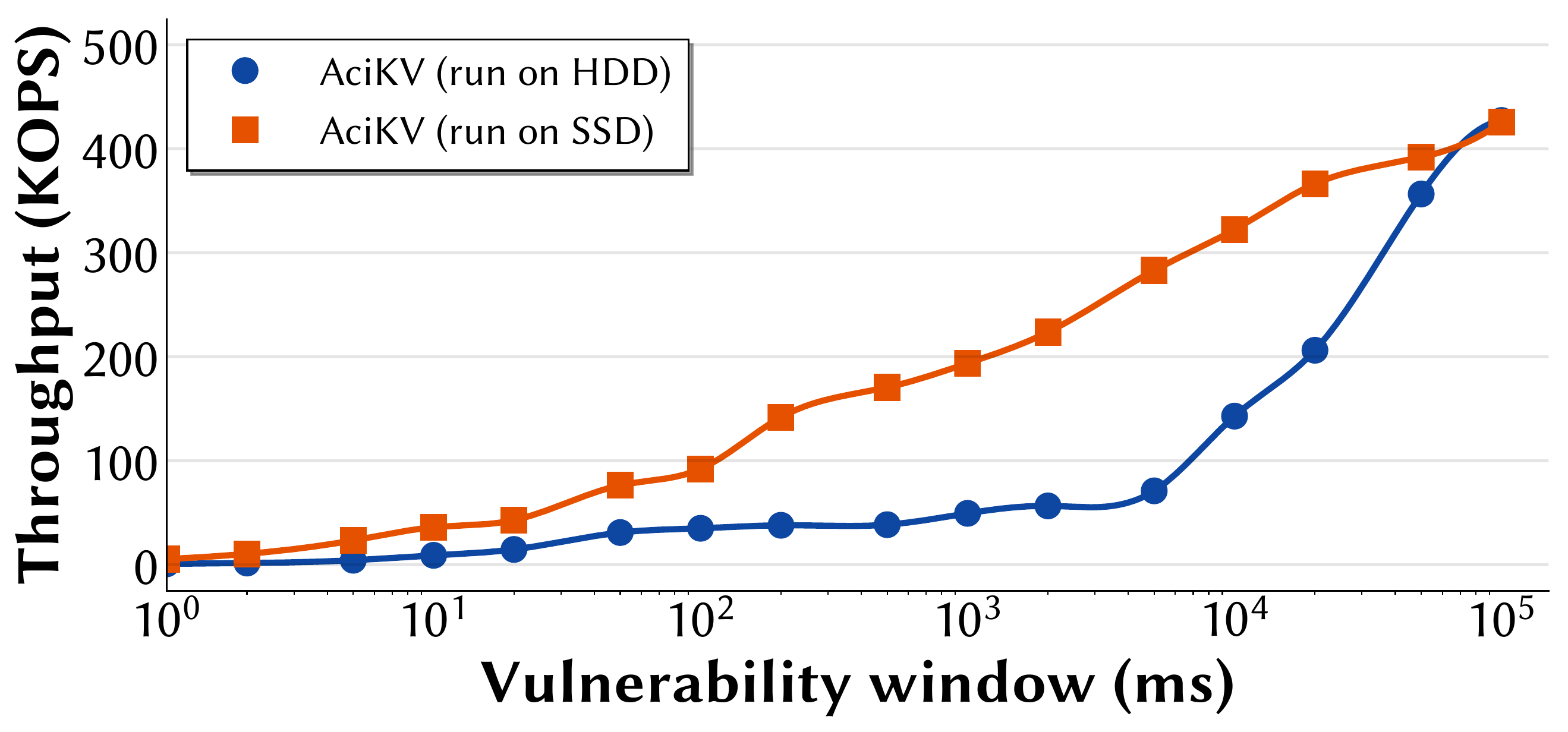}
  \caption{
      Weak durability: trade-off between throughput and vulnerability window.
  }
  \label{graph:ycsb-persist}
\end{figure}

We begin with the configuration of weak durability.
We employed a dedicated thread to invoke \syspersist for every time interval~$k$, which we refer to as the \emph{vulnerability window} to reflect that only transactions committing within the most recent $k$-interval might disappear in the face of a crash.
We used the write-only workload, and tuned the value of~$k$ from 1ms to 100s.
To understand \sys's performance on slow storage media, we also used an HDD.

\autoref{graph:ycsb-persist} shows the throughput with respect to each~$k$.
As expected, a larger vulnerability window gives rise to a higher throughput.
The results clearly indicate that durability requirements have a strong implication for performance.
This experiment also demonstrates \sys's flexibility compared with strong durability---users can pick a suitable value of~$k$ for their application by replaying a similar experiment, perhaps with their own workloads.
For slower media like HDDs, the throughput also increases at a slower rate.
This means that to achieve the same target performance, faster media can have a smaller vulnerability window.

\begin{figure}[t]
  \centering
  \includegraphics[width=\columnwidth]{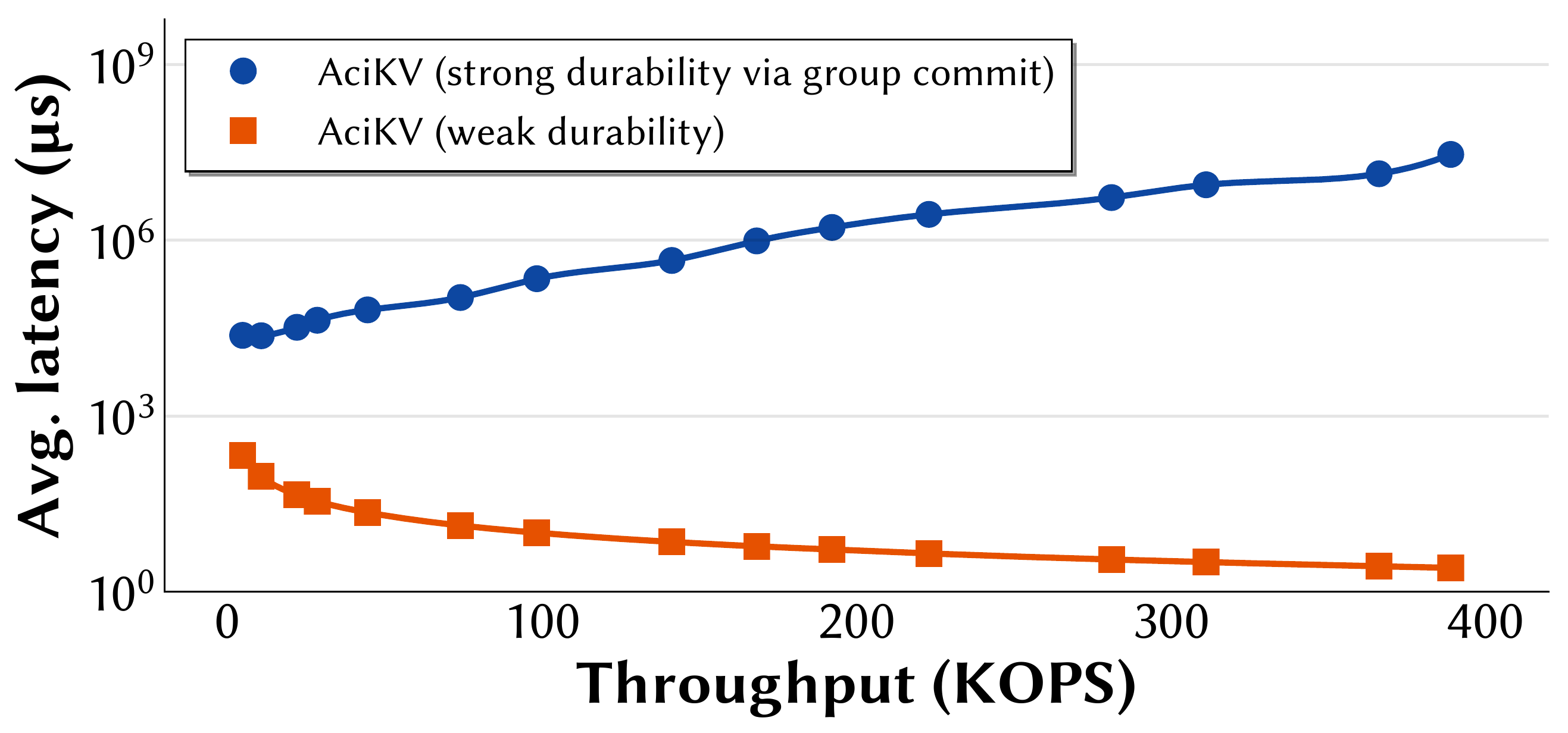}
  \caption{
      Group commit: trade-off between throughput and latency.
  }
  \label{graph:ycsb-group}
\end{figure}

\begin{figure*}[t]
  \centering
  \scalebox{.35}{
    \includegraphics{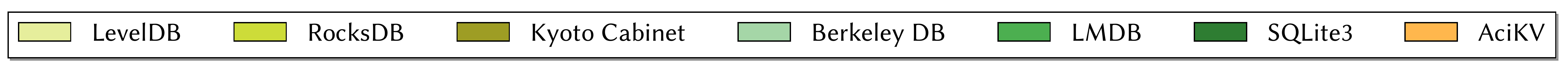}
  }

  \subfigure[Cache size = 8 GB.]{
    \includegraphics[width=\columnwidth]{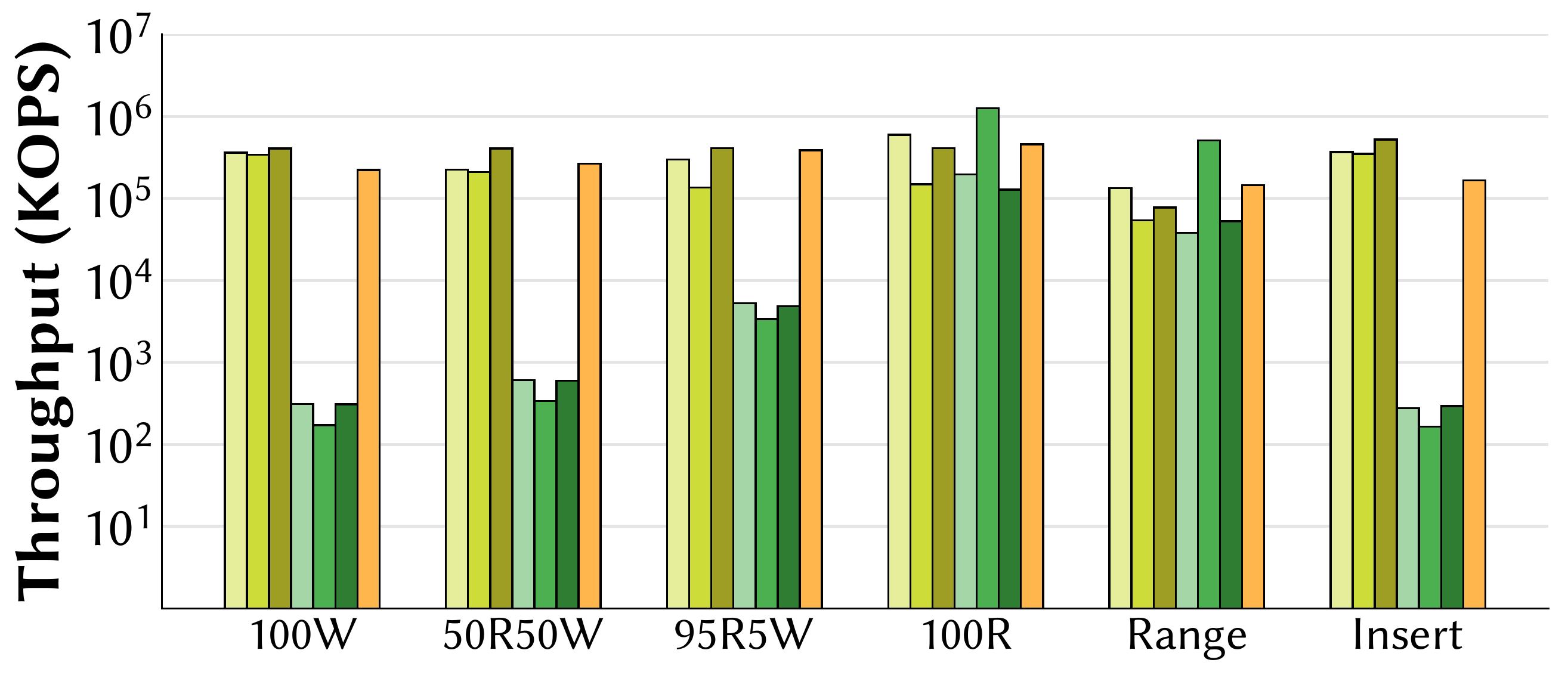}
    \label{graph:ycsb-large}
  }
  \subfigure[Cache size = 4 MB.]{
    \includegraphics[width=\columnwidth]{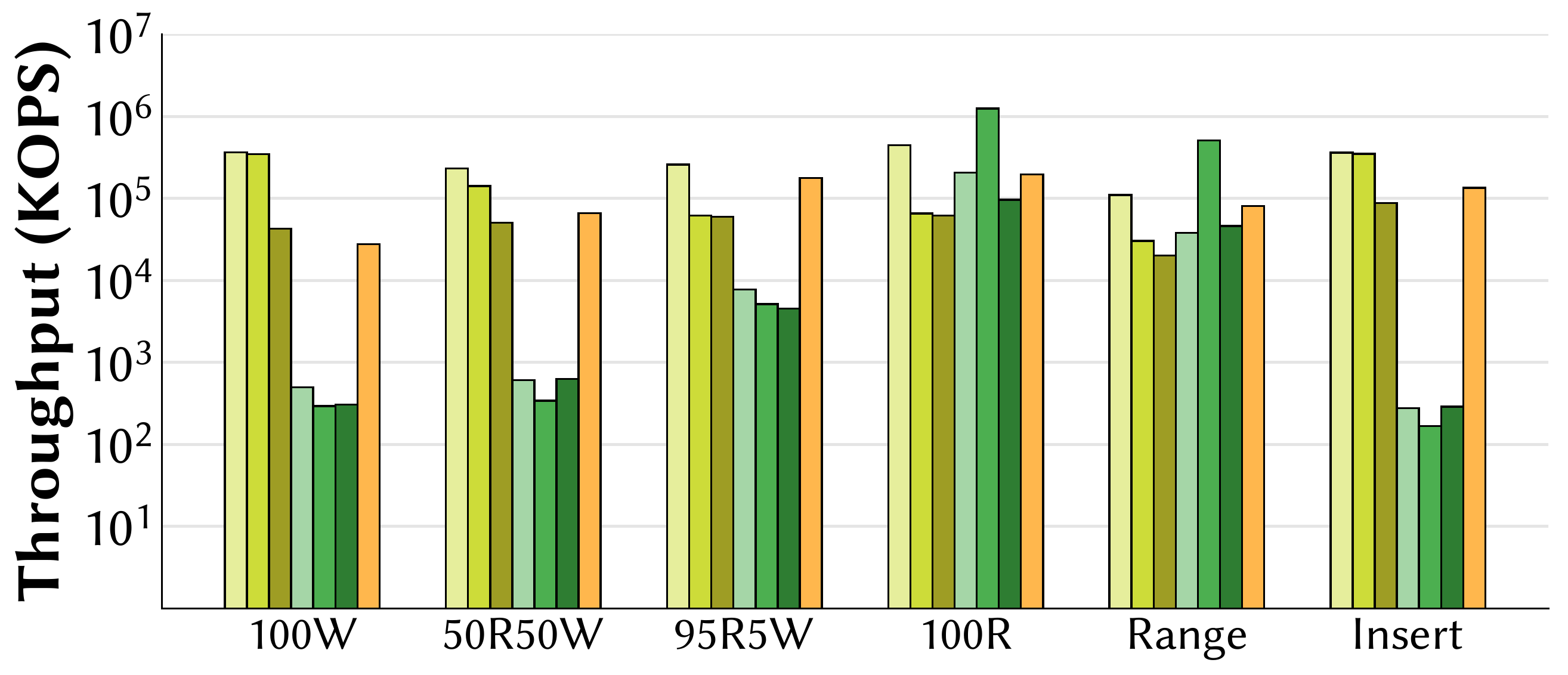}
    \label{graph:ycsb-small}
  }
  \caption{
      YCSB results.
      For \leveldb and \rocksdb, we configured their write buffer size to 120 MB.
      For \sqlite, we set its journal mode to WAL.
      For all systems, we configured them in read-only mode for read-only workload.
  }
  \label{graph:ycsb}
\end{figure*}

Next we study the trade-off imposed by group commit.
Similarly to the prior setting, we tuned the value of $k$ from 1ms to 100s, but this time we reported the average latency against the throughput.
We count the latency of weak durability (the orange line) as the time interval between a \sysbegin-\syscommit pair, and the latency of group commit (the blue line) as the previous latency plus the waiting time, defined as the time interval between the \syscommit and the next \syspersist it encounters.

\autoref{graph:ycsb-group} shows the results, indicating a clear trend: In the group commit case, the latency increases as the throughput increases, whereas in the weak durability case, the latency decreases as the throughput increases.
This shows that group commit forces users to choose between a high throughput or a low latency.
The latency difference between the two techniques is more obvious when aiming for a high throughput.
For instance, in our experiment, for a target throughput of 300 KOPS, the latency of group commit is five orders of magnitude higher than that of weak durability.
The merit of group commit is, of course, strong durability, so it has no vulnerability window.

In summary, with weak durability, users should choose an acceptable vulnerability window for a desired performance.
With group commit, users are required to find a balance between throughput and latency.
For all subsequent experiments, we set the vulnerability window of \sys to 5s.

\subsection{Comparing \sys with existing systems}
\label{subsec:ycsb}

The main goal of \sys is to guarantee transactional properties while maintaining a similar performance to non-transactional systems.
Thus, we compare \sys with six popular disk-based database systems, including three non-transactional systems\footnote{\rocksdb and \kyoto also provide some transactional supports, but our experiment did not use them.}: \leveldb, \rocksdb, and \kyoto, and three transactional systems: \bdb, \lmdb, and \sqlite.
\autoref{tbl:dbinfo} summarizes some key design choices of these systems:

\begin{table}[h]
  \small\centering
  \begin{tabular}{@{} lrrr @{}}
	\toprule
    \textbf{System} & \textbf{Interface} & \textbf{Index} & \textbf{Recovery} \\
	\midrule
    \leveldb & Non-txn KV & LSM-tree & WAL \\
    \rocksdb & Non-txn KV & LSM-tree & WAL \\
    \kyoto & Non-txn KV & B+-tree & WAL + SP \\
    \bdb & Txn KV & B+-tree & WAL \\
    \lmdb & Txn KV & B+-tree & CoW \\
    \sqlite & SQL & B+-tree & WAL \\
	\bottomrule
  \end{tabular}
  \caption{Design choices of database systems that are used in our experiment. WAL: write-ahead logging; SP: shadow paging; CoW: copy-on-write.}
  \label{tbl:dbinfo}
\end{table}

We would like to emphasize that every system comes with its own design choice and tuning parameters, and without a total understanding of them, it is hard to make a completely fair comparison.
We therefore ask the reader to ignore relatively small performance difference, and focus more on the order of magnitude.

\autoref{graph:ycsb-large} shows the throughput of each system under different workloads.
The first four bar groups are the results of running the read-or-write workload with read ratio set to 0, 0.5, 0.95, and 1, respectively.
When only writes are issued, the throughput of \sys is 662x--1219x higher than that of the three transactional systems, and is within half of the non-transactional ones.
The performance difference between \sys and other transactional systems decreases as the read ratio increases: with the read ratio set to 0.5 and 0.95, the difference drops to 282x--763x and 52x--111x, respectively.
When only reads are issued, durability is no longer a factor that can affect performance, and thus the results are roughly at the same order.
This is also true for the range-query workload (the 5\textsuperscript{th} bar groups).
For the insertion workload (the 6\textsuperscript{th} bar groups), \sys is 454x--789x faster than the transactional systems, and at worst 4x slower than the non-transactional ones.
The difference between the performance of \sys's updates (209 KOPS) and insertions (140 KOPS) can be mainly attributed to the merging of the skip list and the B+-tree.

The previous experiment uses a large cache (8 GB) that can accommodate the entire database, so disk I/Os are mostly writes for durability.
To understand how a smaller cache can affect the performance, we set the cache size to 4 MB (except for \lmdb, which does not maintain its own cache, and instead relies on the page cache of file systems) and repeated the previous experiment.

\begin{figure*}[t]
  \centering
  \scalebox{.35}{
    \includegraphics{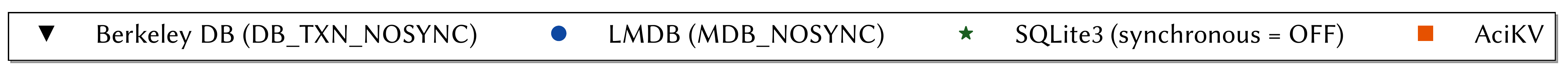}
  }

  \subfigure[100R.]{
    \includegraphics[width=.32\linewidth]{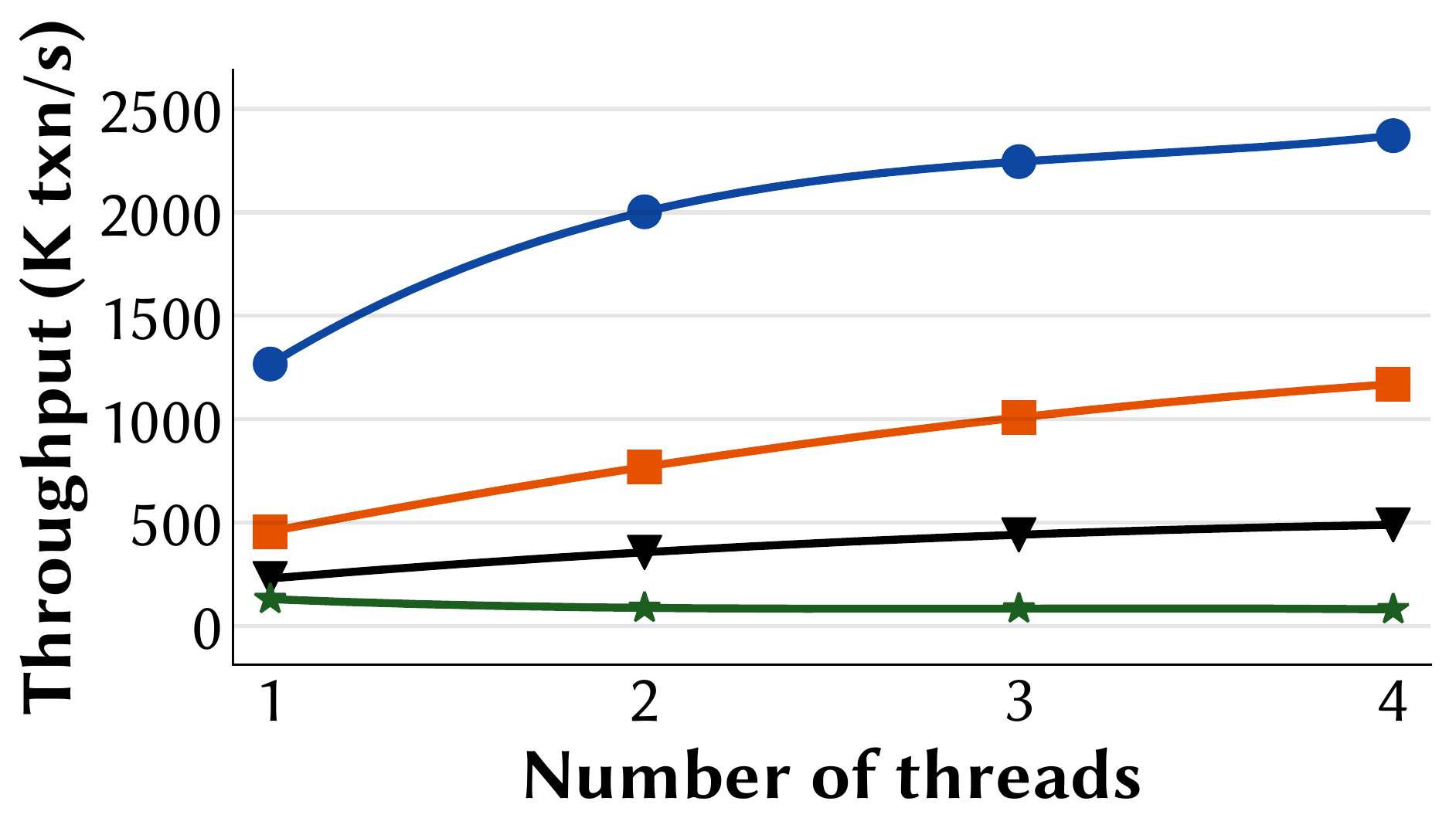}
    \label{graph:ycsb-mt-c}
  }
  \subfigure[Range query.]{
    \includegraphics[width=.32\linewidth]{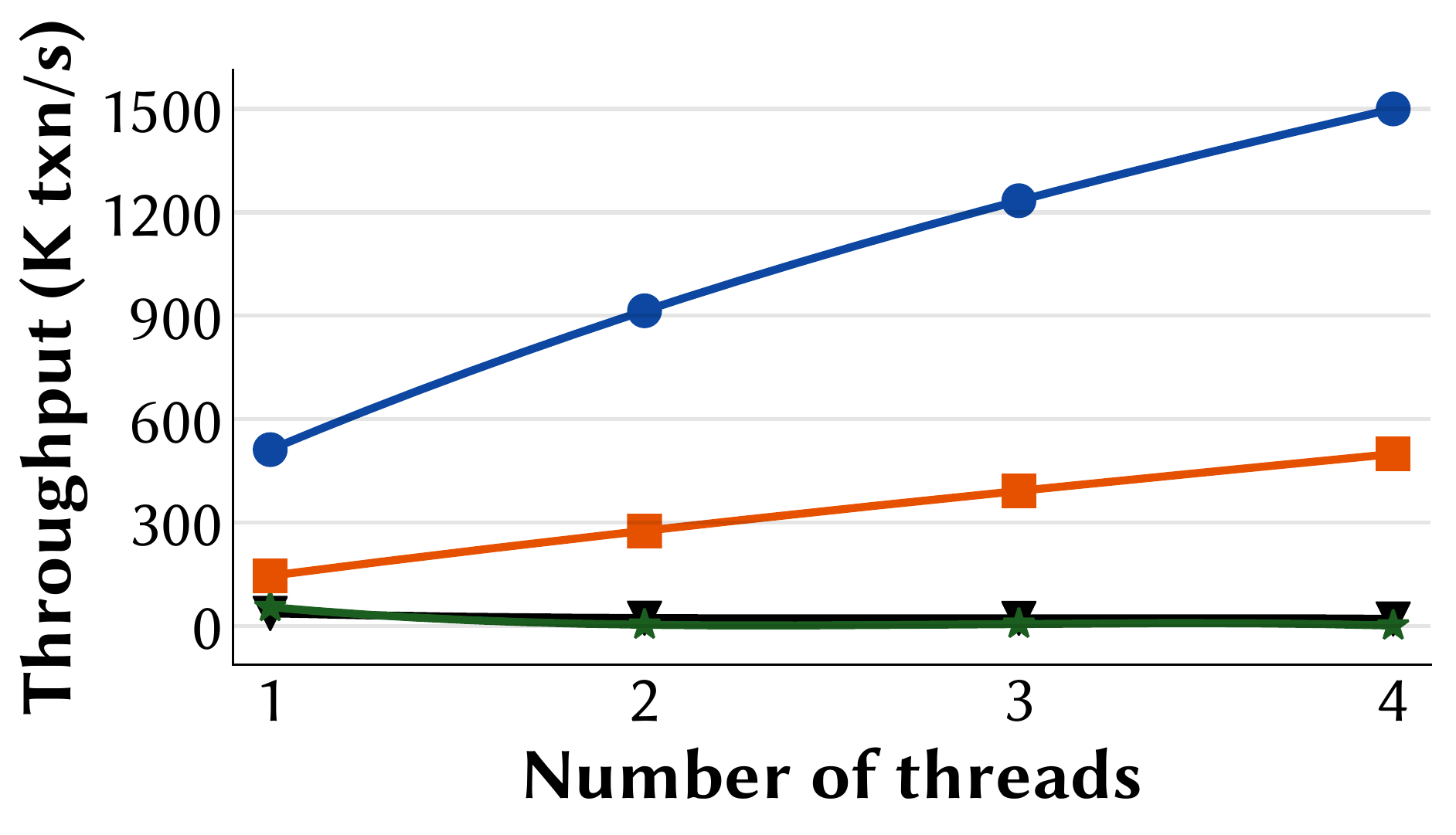}
    \label{graph:ycsb-mt-e}
  }
  \subfigure[95R5W.]{
    \includegraphics[width=.32\linewidth]{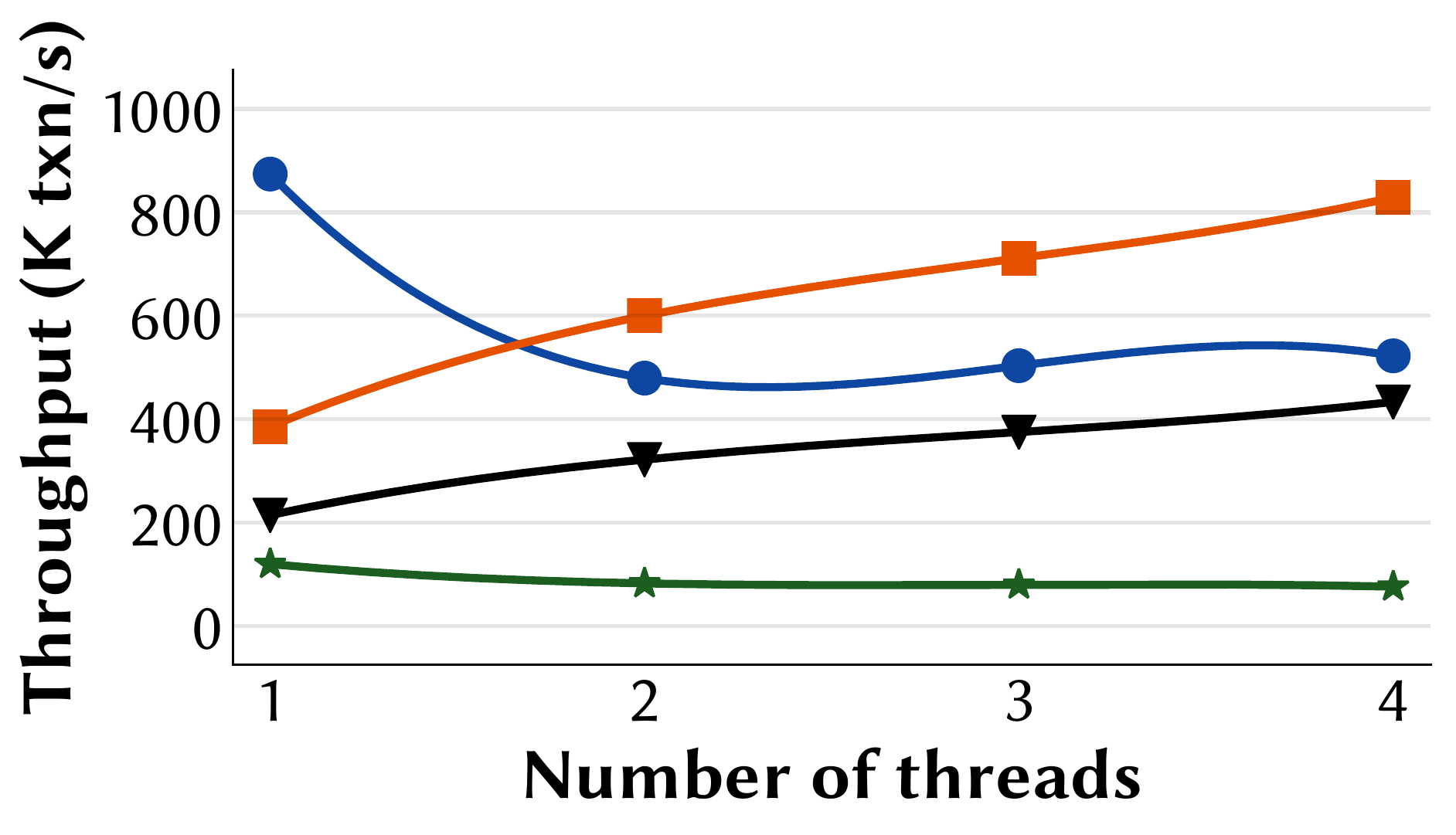}
    \label{graph:ycsb-mt-b}
  }
  \subfigure[Read-modify-write.]{
    \includegraphics[width=.32\linewidth]{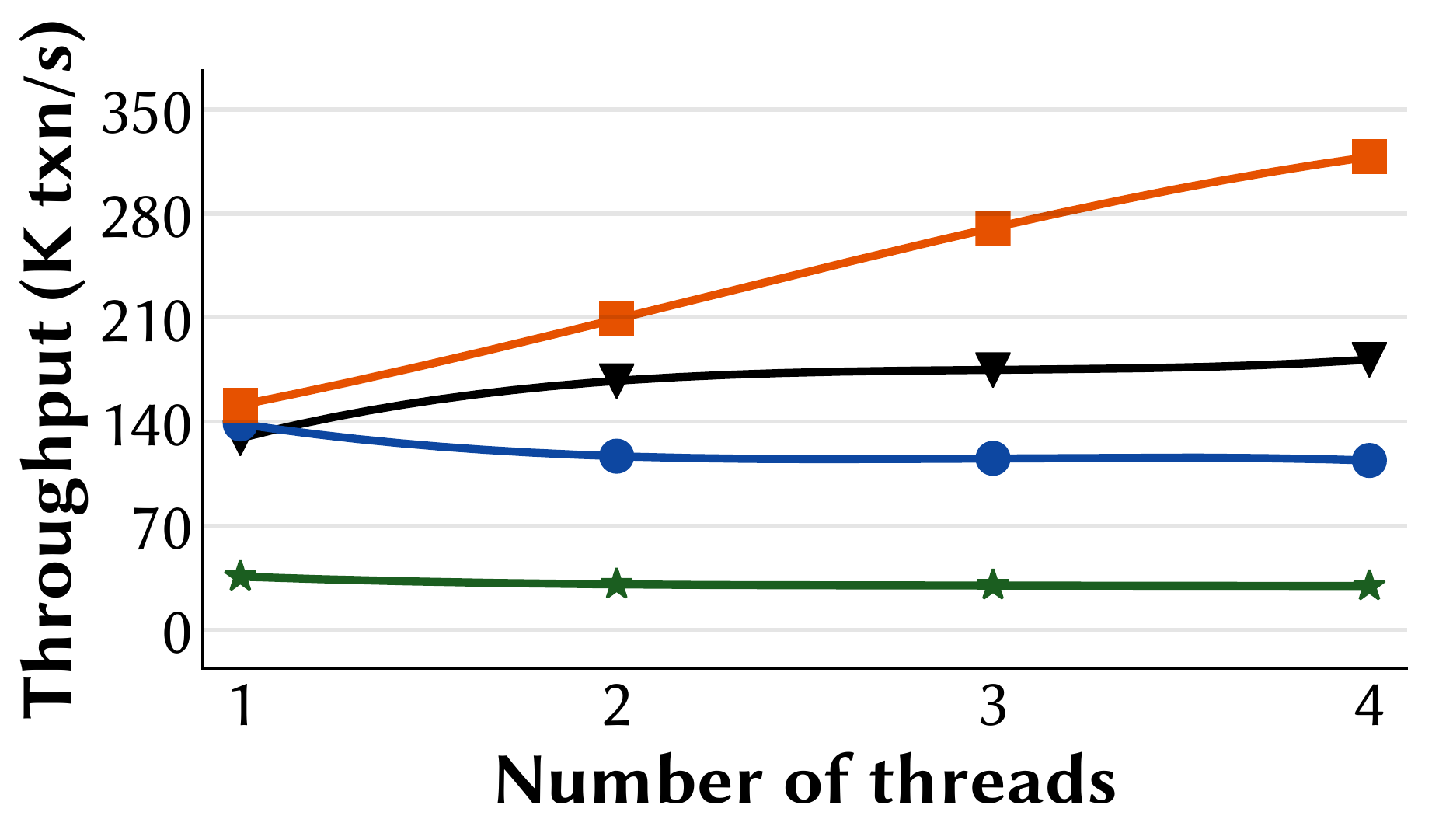}
    \label{graph:ycsb-mt-f}
  }
  \subfigure[100W.]{
    \includegraphics[width=.32\linewidth]{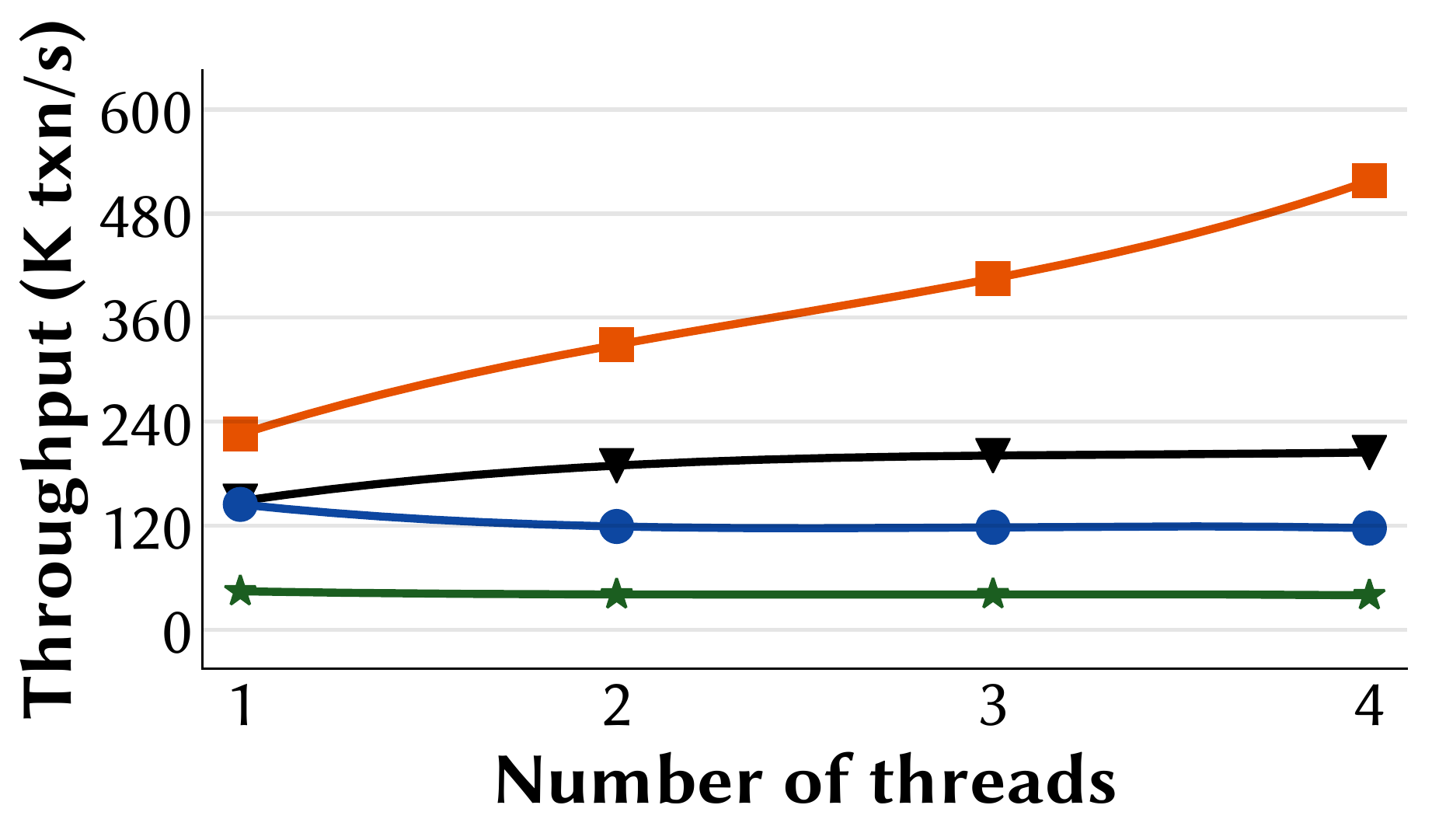}
    \label{graph:ycsb-mt-g}
  }
  \subfigure[Insert.]{
    \includegraphics[width=.32\linewidth]{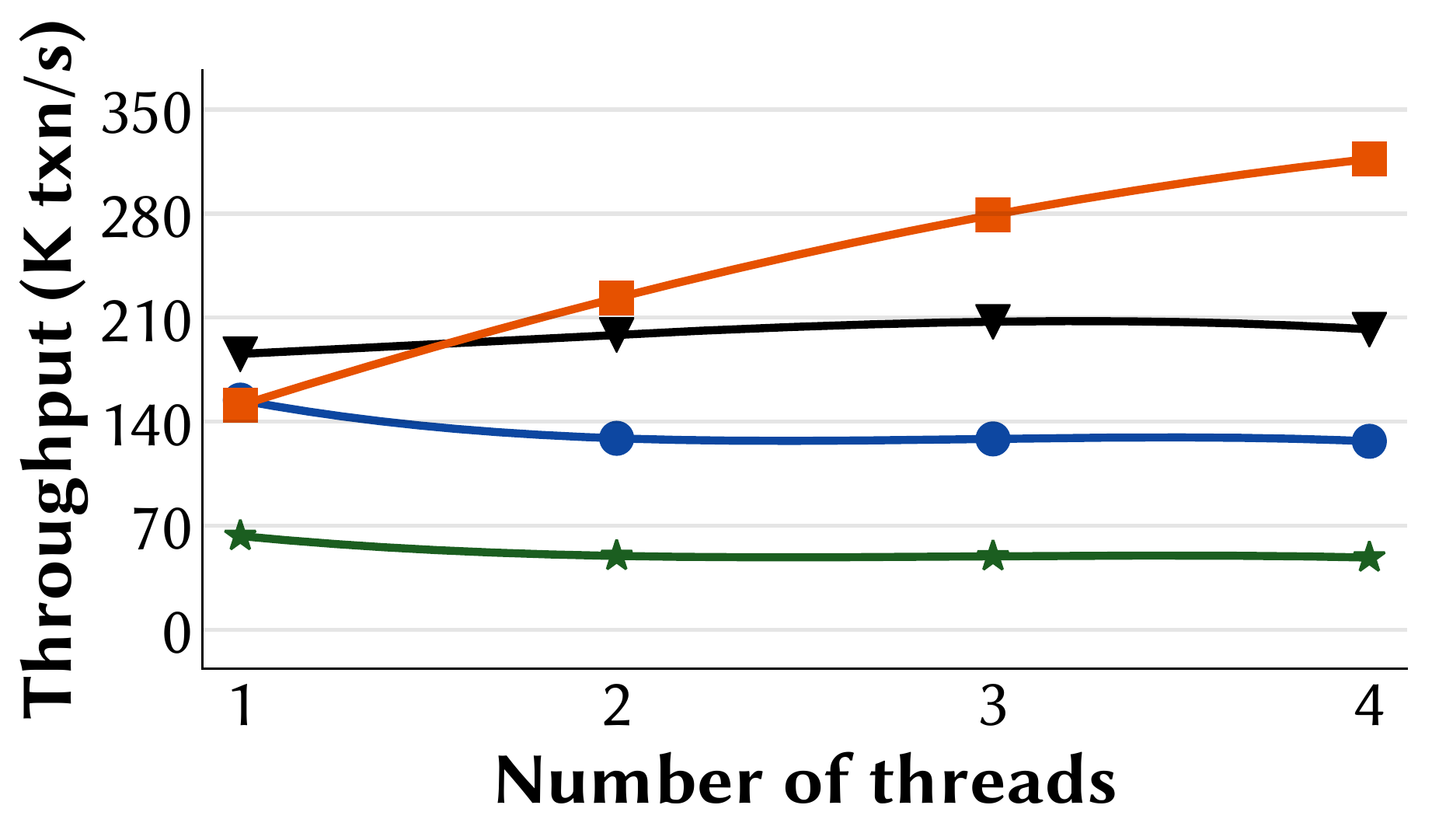}
    \label{graph:ycsb-mt-h}
  }
  \caption{
      Multi-thread YCSB results.
  }
  \label{graph:ycsb-mt}
\end{figure*}

\autoref{graph:ycsb-small} shows the results.
Our analysis is divided into three parts based on the workload types:
(i) For read-only workloads (the 4\textsuperscript{th} and 5\textsuperscript{th} bar groups), a smaller cache degrades the performance of all systems by 25\% to 85\% (\sys decreases by 58\%).
(ii) For write-only workloads (1\textsuperscript{st} and 6\textsuperscript{th} bar groups), all systems except for \sys and \kyoto are not affected by a smaller cache; however, we believe the reason for this is quite different for each system.
For \leveldb and \rocksdb, we suspect the reason lies in the append-only nature of their LSM-tree design, which does not require reading a page before writing a record, and hence no cache misses are triggered; for \bdb and \sqlite, the reason is that the synchronization overhead due to strong durability is the primary bottleneck affecting the performance, so a few reads due to cache misses are almost negligible.
\sys and \kyoto has lower performance with a smaller cache, which is about 10\% of the performance when the cache can accommodate the entire database.
The reason is that \sys and \kyoto need to, before writing a record, read the page containing the record, and thus can be affected by cache misses.
But even with a smaller cache, \sys still performs 90x--166x and 355x--671x better than the other transactional systems under update and insertion workloads, respectively.
(iii) For read-write mix workloads (2\textsuperscript{nd} and 3\textsuperscript{rd} bar groups), the performance degradation due to cache misses is in between read-only and write-only workloads.

In summary, under workloads that involve writes, \sys provides transactional guarantees while achieving similar performance to systems without transactional support.
With a smaller cache, the throughput of \sys drops because of cache misses, but is still much higher than that of transactional systems with a strong durability guarantee.

\subsection{Multicore scalability}
\label{subsec:multithreading}

We built \sys around weak durability.
To understand the benefit of doing so, we compare \sys to other transactional systems that consider strong durability as their ``first-class'' property, and then provide ad-hoc options to relax durability.
As discussed in \autoref{sec:intro} and in \autoref{subsec:txn-sys}, enabling these options is subject to internal and external inconsistency in the event of a crash.
This experiment focuses on the performance aspect, especially \emph{multicore scalability}.

In this experiment, we chose the transactional systems \bdb, \lmdb, and \sqlite, and opened their own weakly durable option.
We used the read-or-write, insertion, range query, and read-modify-write workloads to compare \sys with the above systems.
We changed the number of threads from 1 to 4, and reported the throughput.

Again, we divide our analysis based on workload types.
\autoref{graph:ycsb-mt-c} and \autoref{graph:ycsb-mt-e} show the results of read-only workloads.
Both \sys and \lmdb can scale well under these workloads.
In particular, \lmdb shows great read performance, partly because it is intentionally designed to be a read-optimized database system; for instance, its multi-version concurrency control mechanism frees its readers from acquiring any lock during execution.
The throughput of \lmdb, however, drops when the workloads involve some writes, as shown in \autoref{graph:ycsb-mt-b} and \autoref{graph:ycsb-mt-f}.
For write-only workloads, as shown in \autoref{graph:ycsb-mt-g} and \autoref{graph:ycsb-mt-h}, \sys, \bdb, and \lmdb have similar performance with a single thread, but \sys performs better than the others when there are multiple threads concurrently writing the database.
The primary reason why \sys can scale under these workloads lies in its two-level latch-free design---by absorbing insertions with a concurrent latch-free skip list, transactions do not need to acquire any index latch when traversing the B+-tree.

In summary, \sys is the only system that consistently scales across a wide range of workloads.
Building a system around weak durability not only prevents inconsistency, but also has the potential to discover a better design.

\subsection{Recovery time}
\label{subsec:recovery-time}

\sys relies on shadow paging to ensure crash safety (\autoref{subsec:shadow}).
Unlike logging-based techniques whose recovery time usually depends on \emph{when} a crash occurs (\eg the recovery time for crashes appearing at the end of the log is likely to be longer than that at the start of the log), the shadow paging technique \sys uses has a nearly constant recovery time regardless of when a crash occurs.
Instead, the recovery time depends only on the size of the database.
This means that it is easier to measure \sys's recovery time.

\begin{figure}[t]
  \centering
  \includegraphics[width=\linewidth]{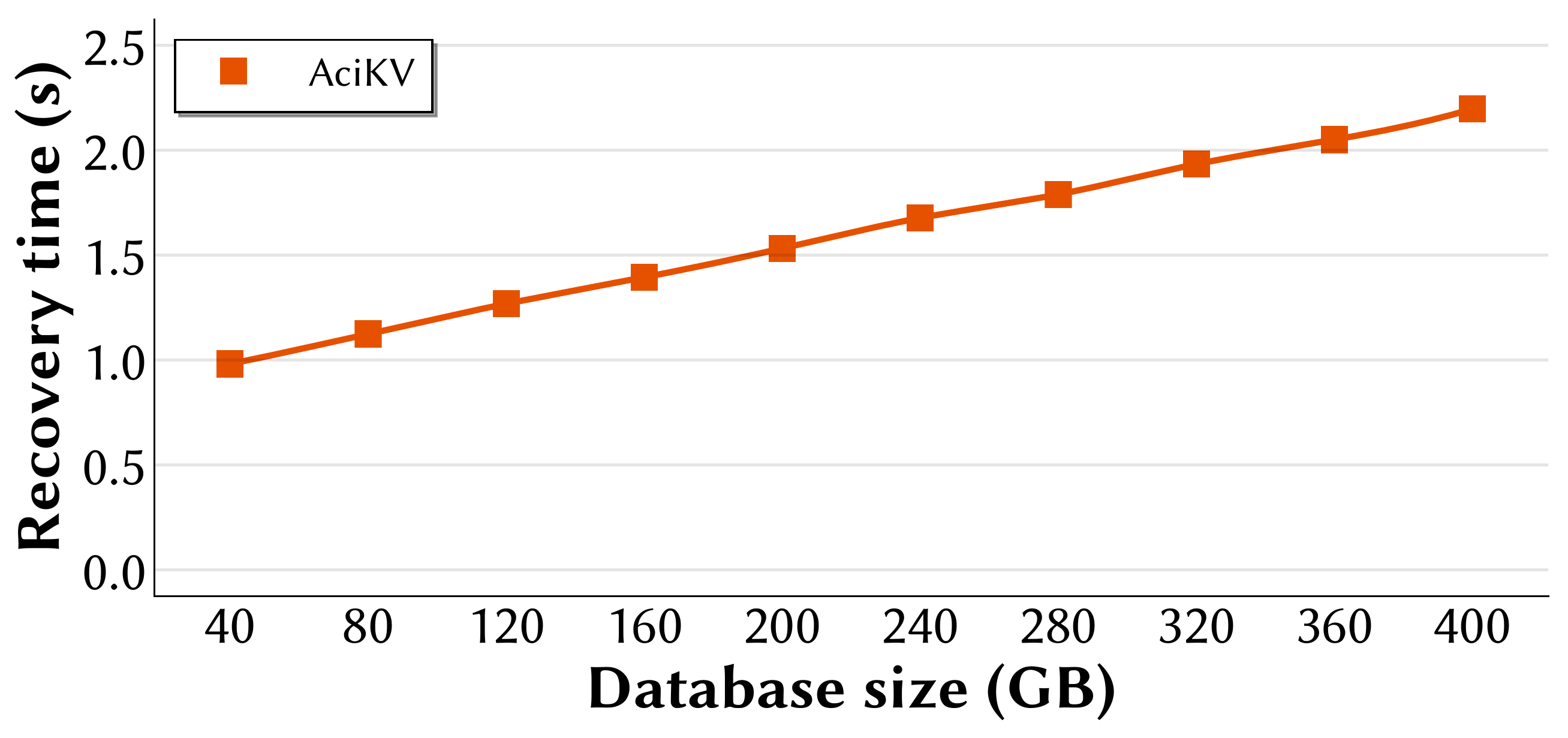}
  \caption{
      Recovery time with respect to different database sizes.
  }
  \label{graph:recovery}
\end{figure}

\autoref{graph:recovery} shows the recovery time of \sys with respect to different database sizes.
The recovery time increases linearly with the database size.
For a 400GB database, the recovery time is less than 3 seconds, which we believe is acceptable in most cases.

\subsection{Memory overhead}
\label{subsec:memory-overhead}

There are two main sources of memory overhead in \sys: one is the in-memory skip list (\autoref{subsec:index}) and the other is the page table used in shadow (\autoref{subsec:shadow}).

We first discuss how we determine the size of the skip list, which depends on two factors: the maximum insertion throughput and the size of each record.
For instance, we allocated 10M skip list entries (each of which can accommodate one record) in the previous experiments; since the skip list is reset every time a \syspersist is issued, and our persist interval is set to 5s, the maximum insertion throughput is thus 2M insertions per second.
Given each record is about 120B, the memory overhead of the skip list is about 1.2 GB.

For the page table used in shadow, the memory overhead is roughly $1/1000$ of the database size, as each 4B table entry maps a 4KB logical page to a physical one (which is also 4KB).
There are two ways to reduce this overhead: we can use a larger page granularity, or we can store the page table on the database file and implement a caching mechanism for the page table.



\section{Related work}
\label{sec:related-work}

\paragraph{High-performance transactional systems.}
Some systems are able to provide full ACID guarantees while maintaining a good performance~\cite{tu13:silo, zheng14:silo-recovery, stonebraker07:hstore, diaconu13:hekaton, dragojevic15:farm, corbett12:spanner, taft20:cockroachdb}.
While many of these are main-memory database systems, to sustain a high performance while ensuring durability, they still need an efficient way to cope with failures.
For instance, Silo~\cite{tu13:silo, zheng14:silo-recovery} and Hekaton~\cite{diaconu13:hekaton} adopt group commit to amortize the disk latency; FaRM~\cite{dragojevic15:farm} relies on NVM; H-Store~\cite{stonebraker07:hstore}, Spanner~\cite{corbett12:spanner}, and CockroachDB~\cite{taft20:cockroachdb} failover to a replica upon system failures.

This work approaches the goal of retaining both the transaction abstraction and a good performance from a different angle, that is, by relaxing the durability requirement.
Unlike prior systems that are often deployed as a cloud service, \sys can be easily hosted on commodity hardware.

%
%

\paragraph{Alleviating synchronization overhead.}

Lower down the storage stack, file systems are also aware of the performance implication of durability requirements.
OptFS~\cite{chidambaram13:optfs}, BarrierFS~\cite{won18:barrierfs}, Featherstitch~\cite{frost07:featherstitch}, and Horae~\cite{liao20:write} propose to add lightweight ordering primitives to the file system interface, so that applications can use \fsync only when they actually require durability.
NoFS~\cite{chidambaram12:nofs} proposes a novel technique to fully remove the need to enforce ordering constraints on writes, but it cannot implement atomic operations.
Xsyncfs~\cite{nightingale06:xsyncfs} adopts a user-centric approach: a write is guaranteed to be durable when an external observer sees the write.
This approach creates for users the illusion that writes are synchronous and keeps the efficiency of asynchronous writes.

\paragraph{ACID at lower layers.}

ACID is not a unique concept in database systems; some file systems and storage devices also provide full or partial support to ACID.
Examples include TxFS~\cite{hu18:txfs}, Valor~\cite{spillane09:valor}, Mime~\cite{chao92:mime}, the Logical Disk~\cite{jonge93:logical-disk}, Stasis~\cite{sears06:stasis}, TxFlash~\cite{prabhakaran08:txflash}, Beyond Block IO~\cite{ouyang11:beyond-bio}, Mars~\cite{coburn13:mars}, LightTx~\cite{lu13:lighttx}, X-FTL~\cite{kang13:xftl}, and Isotope~\cite{shin16:isotope}.

While these systems have been shown to improve the design of their upper layers (in terms of simplicity and efficiency), one downside of this approach is that it often requires substantial changes to the existing software stack.
\sys shares a similar goal to these systems, that is, providing a useful abstraction to relieve the programming burden of users, but targets a higher abstraction level.

\paragraph{Weak durability in the field.}
Many storage systems are by default weakly durable.
For instance, almost none of the major file systems guarantee that a \texttt{write} is made durable unless users explicitly issue an \texttt{fsync}.
While weak durability substantially improves the performance of these systems, it also has some undesirable consequences: notably, (i) complex specification of how the systems should behave on a crash, and (ii) an enormous space of post-crash states, which in turn lead to many crash safety bugs~\cite{zheng14:torturing-db, pillai14:ccbug, kim19:hydra, mohan18:b3}.
As a result, recent research has proposed to reduce the number of post-crash states resulting from weak durability with some ordering guarantees~\cite{chang19:optr, pillai17:ccfs}, to formally specify crash behavior~\cite{bornholt16:ccmodel, ridge15:sibylfs, kokologiannakis21:ext4-persist}, or to formally verify crash safety~\cite{chen15:fscq, amani16:cogent, chen17:dfscq, chajed19:perennial, chajed19:argosy, hance20:veribetrkv, chang20:scftl} of their systems.

\sys's weak durability is much more benign, in that it is complemented with other transactional guarantees.
For instance, users can expect that their transactions are atomic with respect to crashes, and that their integrity constraints will not be accidentally violated in the face of a crash.


%

\section{Conclusion}
\label{s:concl}

We introduce the notion of weakly durable transactions (\acid) that aims to prevent isolation and consistency anomalies, while benefit from the high I/O parallelism granted by modern storage devices.
We discuss, from a theoretical point of view, how to achieve \acid, and use \acid as the new requirement to build \sys.
We conduct extensive experiments to evaluate \sys; results show that \sys can outperform existing transactional systems by orders of magnitude.
We believe \sys represents a useful middle ground between systems that provide high throughput but have few guarantees, and systems that expose the useful transaction abstraction but suffer from high synchronization overhead.

\newpage
\bibliographystyle{abbrvnat}
\bibliography{p,website}

\end{document}